%% file: mainfile_icml17_cameraready.tex
\documentclass[]{article} 

\usepackage[accepted,nohyperref]{icml2017} 

\usepackage{algorithmic}
\usepackage{wrapfig}
\input{defs}

\allowdisplaybreaks

\usepackage{enumitem}
\setlist[itemize]{leftmargin=6mm}

\icmltitlerunning{Guarantees for Greedy  Maximization of Non-submodular Functions with Applications}
\graphicspath{{./figs/}}

\begin{document}

\twocolumn[
\icmltitle{
Guarantees for Greedy  Maximization
of\\ Non-submodular Functions with Applications}

%\icmlauthor{Andrew An Bian}{ybian@inf.ethz.ch}
%\icmlauthor{Joachim M. Buhmann}{jbuhmann@inf.ethz.ch}
%\icmlauthor{Andreas Krause}{krausea@ethz.ch}
%\icmlauthor{Sebastian Tschiatschek}{ sebastian.tschiatschek@inf.ethz.ch}
%\icmladdress{Department of Computer Science, ETH Zurich}
\icmlsetsymbol{equal}{*}
\icmlsetsymbol{note}{\color{blue}$\dagger$}

\begin{icmlauthorlist}
\icmlauthor{Andrew  An Bian}{note,ed}
\icmlauthor{Joachim M. Buhmann}{ed}
\icmlauthor{Andreas Krause}{ed}
\icmlauthor{Sebastian Tschiatschek}{ed}
\end{icmlauthorlist}

\icmlaffiliation{ed}{Department of Computer Science, ETH Zurich,  Zurich, Switzerland}
%\icmlaffiliation{yab}{{\color{blue}{Now known as (name written as) Yatao A. Bian. ORCID iD: \href{https://orcid.org/0000-0002-2368-4084}{orcid.org/0000-0002-2368-4084}}}}

% , Zurich, Switzerland

\icmlcorrespondingauthor{Joachim M. Buhmann}{{jbuhmann@inf.ethz.ch}}
\icmlcorrespondingauthor{Andreas Krause}{{krausea@ethz.ch}}
% You may provide any keywords that you 
% find helpful for describing your paper; these are used to populate 
% the "keywords" metadata in the PDF but will not be shown in the document
\icmlkeywords{Greedy algorithm, non-submodular, ICML}
\vskip 0.3in
]

% this must go after the closing bracket ] following \twocolumn[ ...

% This command actually creates the footnote in the first column
% listing the affiliations and the copyright notice.
% The command takes one argument, which is text to display at the start of the footnote.
% The \icmlEqualContribution command is standard text for equal contribution.
% Remove it (just {}) if you do not need this facility.

%\printAffiliationsAndNotice{}  % leave blank if no need to mention equal contribution
\printAffiliationsAndNotice{{\color{blue}{$^\dagger$Now known as Yatao A. Bian $<$ybian@inf.ethz.ch$>$. ORCID: \href{https://orcid.org/0000-0002-2368-4084}{orcid.org/0000-0002-2368-4084}}}} % otherwise use the standard text.

%\printAffiliationsAndNotice{} % otherwise use the standard text.

\input{abs.tex}

\input{introduction.tex}

\input{general_ratio_curvature.tex}

\input{guarantee_proof.tex}

\input{app}

\input{exp.tex}

\input{related.tex}

\input{disc.tex}

\subsubsection*{Acknowledgements}

%\textsc{Acknowledgements}

The authors would like to thank  Adish Singla, Kfir Y. Levy   and Aurelien Lucchi   for valuable discussions. 
This research was partially supported by  ERC StG 307036 and the  
Max Planck ETH Center for Learning Systems.
This work was done in part while Andreas Krause was visiting the Simons Institute for the Theory of Computing. 

\clearpage
\bibliographystyle{icml2017}
{
%\small 
\bibliography{bib}
}

\clearpage
\appendix
\input{appendix}

\end{document}

%% file: defs.tex
\usepackage[usenames]{xcolor}
\usepackage[]{hyperref}
\usepackage{url}
\definecolor{cite_color}{RGB}{0, 51, 153}
\definecolor{link_color}{RGB}{153, 0,0}  %  red
\definecolor{url_color}{RGB}{153, 102,  0}
\definecolor{emp_color}{RGB}{0,0,255}
\hypersetup{
 colorlinks=true,
 citecolor=cite_color,
 linkcolor=link_color,
 urlcolor=url_color}

\usepackage{times}
\usepackage{amsmath}
\usepackage{amsthm}
\usepackage{amssymb}
\usepackage{graphicx}
\usepackage{subfig}
\usepackage{multirow}
\usepackage{booktabs}
\usepackage{mathrsfs}
\usepackage{wrapfig}
\usepackage{enumerate}
\usepackage{bm}
\usepackage{bbm}
\usepackage{tabularx}
\usepackage{natbib}
\usepackage[vlined, ruled]{algorithm2e}
\SetCommentSty{mycommfont}
\SetKwInOut{Input}{input}
\SetKwInOut{Output}{output}

 \usepackage{cleveref} % should be put after other packages 
 \crefname{section}{Section}{Sections}
 \crefname{theorem}{Theorem}{Theorems}
 \crefname{lemma}{Lemma}{Lemmas}
 \crefname{appendix}{Appendix}{Appendix}
 \crefname{equation}{Eq.}{Equations}
 \crefname{proposition}{Proposition}{Propositions}
 \crefname{claim}{Claim}{Claims}
  \crefname{algorithm}{Alg.}{Algorithms}
 \crefname{figure}{Fig.}{Figures}
 \crefname{table}{Table}{Tables}
  \crefname{remark}{Remark}{Remarks}
   \crefname{definition}{Def.}{Definitions}
\usepackage{mathtools}

\newtheorem{theorem}{Theorem}
\newtheorem{lemma}{Lemma}
\newtheorem{proposition}{Proposition}

\newtheorem{definition}{Definition}
\newtheorem{remark}{Remark}
\newtheorem{claim}{Claim}

\def\C{{\cal C}}

\def\E{{\mathbb E}}

\def\N{{\cal N}}

\def\M{{\cal M}}
\def \I {{\cal{I}}} 
  
\def\opt{\ensuremath{\Omega^*}}

\def \c{\bm{c}}
\def \u{\bm{u}}

\def \w{\bm{w}}

\def \b{\bm{b}}
\def \d{\bm{d}}

\def \x{\bm{x}}
\def \y{\bm{y}}

\def \e{\bm{e}}
\def \bmA{\mathbf{A}}
\def \bmB{\mathbf{B}}
\def \bmX{\mathbf{X}}
\def \bmI{\mathbf{I}}
\def \bmC{\mathbf{C}}
\def \bmD{\mathbf{D}}

\def \bmP{\mathbf{P}}

\def \bmLambda{\mathbf{\Lambda}}
\def \bmSigma{\mathbf{\Sigma}}

\def \bmtheta{\bm{\theta}}
\def \bmlambda{\bm{\lambda}}
\def \bmalpha{\bm{\alpha}}

\def \bmxi{\bm{\xi}}

\def \P{{\cal{P}}}

\def \R{{\mathbb{R}}}
\def \trans{\top}
\newcommand{\tr}[1]{\text{tr}(#1)}

\def \V{{\cal{V}}}
\def \E{{\mathbb{E}}} %  expection

\newcommand{\dom}[1]{{\texttt{dom}(#1)}}
\newcommand{\dep}{{\text{Dep}}} % dependency set of a set function 
\newcommand{\algname}[1]{{\textsc{#1}}}

\newcommand{\spt}[1]{{\text{supp}}(#1)}
\newcommand{\dtp}[2]{\langle #1, #2\rangle}% dot product
\newcommand{\de}[1]{\text{det}(#1)}% determinant of a square matrix

\newcommand{\groundset}{\ensuremath{\mathcal{V}}}

%% file: abs.tex
%!TEX root = main_ratio_curvature_icml_format.tex

\begin{abstract}
We investigate the performance of the standard \algname{Greedy} algorithm 
for cardinality constrained maximization of non-submodular nondecreasing set functions.
While there are strong theoretical guarantees on the performance of \algname{Greedy} for maximizing submodular functions, there are few guarantees for non-submodular ones.
However, \algname{Greedy} enjoys strong empirical performance for  many important non-submodular functions, e.g., the Bayesian A-optimality objective in experimental design.
We prove theoretical guarantees supporting the empirical performance. Our 
guarantees are characterized by a combination of the 
(generalized) \textit{curvature} $\alpha$ and 
 the  \textit{submodularity ratio} $\gamma$.
In particular, we prove that \algname{Greedy}  enjoys a \emph{tight} approximation guarantee of  
 $\frac{1}{\alpha}(1- e^{-\gamma\alpha})$ for cardinality constrained maximization. 
In addition, we bound the submodularity ratio and 
curvature for several important real-world objectives, including the Bayesian
A-optimality objective, the determinantal
function of a square submatrix and certain linear programs with combinatorial constraints. 
We experimentally validate our theoretical findings for both synthetic and  real-world applications.

\end{abstract}

%% file: introduction.tex
%!TEX root = mainfile_icml17_cameraready.tex

\section{Introduction}

%Consider the important problems of \emph{experimental design} and \emph{sparse modeling}. 

Many important 
problems, such as {experimental design} and {sparse modeling},  are naturally formulated as a subset selection problem,  where   a set function $F(S)$ over  a $K$-cardinality constraint is maximized, i.e.,
\begin{align}\label{eq1}
  \max_{S\subseteq  \groundset,  |S| \leq K} F(S), \tag{P}
\end{align}
where  $\groundset =\{v_1, \ldots, v_n\}$ is  the ground set. 
Specifically, in experimental design, the goal is to select a set of experiments to perform such that some statistical criterion is optimized.
%e.g., the variance of certain parameter estimates is minimized. 
This problem arises naturally in domains where performing experiments is costly.
% e.g., the medical domain. 
 In sparse modeling, the task is to identify sparse representations of signals, enabling interpretability and robustness in high-dimensional statistical problems---properties that are crucial in modern data analysis.

%These problems are naturally cast as subset selection problems such that  a set function $F(S)$ over  a $K$-cardinality constraint is maximized, i.e.,
%\begin{align}\label{eq1}
%  \max_{S\subseteq  \groundset,  |S| \leq K} F(S), \tag{P}
%\end{align}
%where  $\groundset =\{v_1, \ldots, v_n\}$ is  the ground set. 
Frequently, the standard \algname{Greedy} algorithm (\cref{alg:greedy}) is used to (approximately) solve \eqref{eq1}.
\begin{algorithm}[htbp]
  \DontPrintSemicolon
%\small
  \caption{The \algname{Greedy} Algorithm}  %  
  \label{alg:greedy}
    \KwIn{ Ground set $\V$, set function $F\colon \! 2^\V \! \rightarrow\! \R_+$, budget $K$}
     {$S^0 \leftarrow \emptyset$\;}
    \For{$t=1,\ldots,K$}{
       {$v^* \leftarrow \arg \max_{v \in \V \setminus S^{t-1}} F(S^{t-1} \cup \{ v\} )- F(S^{t-1})$ \;}
       {$S^{t} \leftarrow S^{t-1} \cup \{v^*\}$\;}
     }
    \KwOut {$S^K$}
\end{algorithm}
%
%The \algname{Greedy} algorithm is a fundamental scheme in 
%combinatorial optimization, it is one of the rare cases where
%an algorithm can be  used to define 
% a combinatorial structure \citep{Korte:2007:COT:1564997}, for instance,  matroid and greedoid. 
For the case that $F(S)$ is a monotone nondecreasing  \emph{submodular} set function\footnote{$F(\cdot)$ is monotone nondecreasing if $\forall A\subseteq \groundset, v\in \groundset$, $F(A\cup \{v \})\geq F(A)$.  $F(\cdot)$ is submodular   iff it satisfies the  diminishing
returns property  $F(A \cup \{v\}) - F(A) \geq F(B \cup \{v\}) - F(B)$ for all $A \subseteq B \subseteq \groundset \setminus \{v\}$. Assume wlog. that  $F(\cdot)$ is normalized, i.e.,  $F(\emptyset) = 0$.},
the \algname{Greedy} algorithm enjoys the multiplicative approximation guarantee of  $(1-1/e)$~\citep{nemhauser1978analysis,DBLP:conf/stoc/Vondrak08,krause2012submodular}. 
This constant factor can be improved by refining the characterization of the objective using the \textit{curvature}~\citep{conforti1984submodular,vondrak2010submodularity,iyer2013curvature}, which informally quantifies how \emph{close} a submodular function is to being modular (i.e., $F(S)$ and $-F(S)$ are submodular). 

However, for many applications, including  experimental design and sparse Gaussian processes \citep{lawrence2003fast}, $F(S)$ is in general not submodular~\cite{krause2008near} and the above guarantee does not hold.
In practice, however, the standard \algname{Greedy} algorithm often
achieves very good performance on these applications, e.g., in
 subset selection  with 
the $R^2$ (squared multiple correlation) objective~\citep{das2011submodular}. To explain the good empirical performance, \citet{das2011submodular}
 proposed the \emph{submodularity ratio}, a quantity characterizing how \emph{close} a set function is to being submodular. 

Another important class of non-submodular  set functions comes as the auxiliary function when  optimizing  a continuous function $f(\x)$ 
s.t. combinatorial  constraints, i.e.,  $\min_{\x\in \C,  \spt{\x}\in \I} f(\x)$, 
where
$\spt{\x}:= \{i \mid x_i \neq 0\}$ is the support set  of $\x$,  $\C$ is
a convex set, and $\I$ is the independent sets  of the combinatorial structure. 
One of the most popular ways to solve this problem is to 
use the \algname{Greedy} algorithm to maximize the auxiliary function $F(S) \coloneqq \max_{\x\in \C,  \spt{\x}\subseteq S}-f(\x)$. 
This setting covers various important  applications, to name a few, feature selection \citep{guyon2003introduction}, 
sparse approximation \citep{das2008algorithms,krause2010submodular}, 
sparse recovery \citep{candes2006stable},  
sparse M-estimation  \citep{jain2014iterative},  
linear programming (LP) with combinatorial constraints,
and column subset selection \citep{altschuler2016greedy}.
Recently, \citet{elenberg2016restricted} proved that if  $f(\x)$ has $L$-restricted smoothness and $m$-restricted   strong
convexity, then the submodularity ratio of $F(S)$	is lower bounded by $m/L$. 
This result significantly enlarges the domain where the \algname{Greedy}
algorithm can be applied. 
%\sebastian{Should we comment on how this works for the independent sets? (it is not specified currently)}

In this paper, we combine and generalize the ideas of  \emph{curvature} and \emph{submodularity ratio} 
to derive improved constant factor approximation guarantees of the \algname{Greedy} algorithm. Our guarantees allow us
to better characterize the empirical success of applying \algname{Greedy} on
a significantly  larger class of non-submodular functions. Furthermore, we bound these characteristics for 
important applications, rendering the usage of  \algname{Greedy}  a principled choice rather than a mere
heuristic.
Our {main contributions are:}
\begin{itemize}
%\vspace{-.2cm}
\item 
We prove  the  \emph{first tight}  constant-factor  approximation guarantees for 
 \algname{Greedy}  on maximizing 
non-submodular nondecreasing set functions s.t.\ a cardinality constraint,  characterized  by a novel \emph{combination} of the  (generalized) notions  of submodularity ratio $\gamma$ and curvature $\alpha$. 
%Our proof techniques are  
%different from previous proofs of  \algname{Greedy}. 

\item By theoretically bounding parameters ($\gamma, \alpha$) for  
several 
important objectives, including Bayesian A-optimality
in experimental design, the determinantal function of a square submatrix and maximization of LPs with combinatorial constraints, 
our theory implies the \emph{first} guarantees for   them. 
%  Furthermore, we obtain improved approximation ratios for the subset 
%selection problem using the  $R^2$ objective.

\item  Lastly, we experimentally
validate our theory on several real-world applications. 
It is worth noting that for the Bayesian A-optimality objective,   
\algname{Greedy} 
generates comparable solutions as  the classically used semidefinite programming (SDP) based method,
but is usually two orders of magnitude faster. 
\end{itemize}

\textbf{Notation.}
We use boldface letters, e.g., $\x$, to represent
vectors, and capital boldface letters, e.g., $\bmA$, to denote matrices. 
$x_i$ is the $i^{\text{th}}$ entry of the vector $\x$. 
We refer to $\groundset =\{v_1, ..., v_n\}$ as the ground set. 
We use $f(\cdot)$ to denote a continuous function, and $F(\cdot)$ to
represent a set function.  $\spt{\x} \coloneqq \{i\in {\groundset} \;|\; x_i \neq 0\}$ is the support set of the vector $\x$, and $[n] \coloneqq \{1, ..., n\}$ for an integer $n\geq 1$.
We denote the marginal gain of a set $\Omega \subseteq \groundset$ in context of a set $S \subseteq \groundset$ as $\rho_{\Omega}(S) \coloneqq F(\Omega \cup S) - F(S)$. For $v \in \groundset$, we use the shorthand $\rho_v(S)$ for $\rho_{\{v\}}(S)$.

%% file: general_ratio_curvature.tex
%!TEX root = mainfile_icml17_cameraready.tex 

\section{Submodularity Ratio and Curvature}\label{sec_defs}

In this section we provide   the \emph{submodularity ratio} and \emph{curvature} for general, not necessarily  submodular functions\footnote{Curvature is commonly  defined for submodular
functions. \citet{sviridenko2015optimal} presented a notion
of curvature for monotone non-submodular functions. 
We show in  \cref{app_classical_defs} the  
details of   these  notions and the relations to ours. Additionally, we prove in \cref{append_remark} of \cref{append_jan} that our combination of  curvature and submodularity ratio  is more expressive than  that of \citet{sviridenko2015optimal} in characterizing the maximization of  problem~\eqref{eq1} using standard \algname{Greedy}.}, they are natural extensions
of the classical ones. 
Let $S^0=\emptyset$, $S^t = \{j_1, ..., j_t \}, t=1,..., K$
be the successive sets chosen by  \algname{Greedy}. % for maximizing function $F(\cdot)$ with a $K$-cardinality constraint .
For brevity,  let $\rho_t \coloneqq \rho_{j_t}(S^{t-1})$ be the marginal gain of \algname{Greedy} in step $t$.
\begin{definition}[Submodularity ratio \citep{das2011submodular}]\label{def:gen-submod-ratio}
The  \emph{submodularity ratio}
of a non-negative set function $F(\cdot)$ is 
the largest scalar $\gamma$ s.t.
%\begin{flalign}
%\gamma \coloneqq \min_{\Omega}  \min_{S}\frac{\sum_{\omega\in \Omega\backslash S}\rho_{\omega}(S)}{\rho_{\Omega}(S)}
%\end{flalign}
\begin{flalign}\notag 
\sum\nolimits_{\omega\in \Omega\backslash S}\rho_{\omega}(S) \geq \gamma \rho_{\Omega}(S), \forall\; \Omega, S\subseteq \groundset.
\end{flalign}
The \emph{greedy submodularity ratio} is the largest scalar $\gamma^G$ s.t.
%\begin{flalign}
%\gamma^G \coloneqq  \min_{\Omega: |\Omega| = K} \min_{0\leq t\leq K-1}\frac{\sum_{\omega\in \Omega\backslash S^t}\rho_{\omega}(S^t)}{\rho_{\Omega}(S^t)}.
%\end{flalign}
\begin{flalign}\notag 
 {\sum_{\omega\in \Omega\backslash S^t}\!\! \rho_{\omega}(S^t)} \geq \gamma^G{\rho_{\Omega}(S^t)}, \forall  |\Omega| \!=\! K,    t=0,\ldots, K-1. 
\end{flalign}
\end{definition}
%When both the numerator and demoninator are zero, we define $\gamma$ and $\gamma^G$ to be  1, which is consistent with how it is used in \cref{eq_ratio_used} of \cref{lem_1} .
It is easy to see that   $\gamma^G \geq \gamma$. 
The submodularity ratio measures to what extent $F(\cdot)$ has 
submodular properties. We  make the following observations: 
\begin{remark}\label{obs_ratio}
For a nondecreasing function $F(\cdot)$, it holds 
a)  $\gamma, \gamma^G \in [0, 1]$; b) $F(\cdot)$ is submodular iff $\gamma = 1$.
\end{remark}

\begin{definition}[Generalized  curvature]\label{def_cur}
%\citep{conforti1984submodular}
The \emph{curvature}
of a non-negative  function $F(\cdot)$ is the smallest scalar $\alpha$ s.t.
%\begin{flalign}\notag 
%\alpha :=1-\min_{\Omega: |\Omega| = K, S: |S|<K}\min_{i\in S \backslash \Omega} \frac{\rho_{i}(S\setminus \{i\} \cup \Omega)}{\rho_{i}(S\setminus \{i\})} . 
%\end{flalign}
\begin{align}\nonumber 
%  \rho_{i}(S\setminus \{i\} \cup \Omega) \geq (1-\alpha){\rho_{i}(S\setminus \{i\})}, \forall \Omega, S\subseteq \groundset, i\in S \backslash \Omega. 
  & \rho_{i}(S\setminus \{i\} \cup \Omega) \geq (1-\alpha){\rho_{i}(S\setminus \{i\})},\\\notag 
  & \forall\; \Omega, S\subseteq \groundset, i\in S \backslash \Omega. 
\end{align}
The \emph{greedy curvature}  is the smallest scalar $\alpha^G \geq 0$ s.t.
\begin{align}\notag 
 &{\rho_{j_i}(S^{i-1}\cup \Omega)}\geq (1-\alpha^G){\rho_{j_i}(S^{i-1})},\\\notag 
&   \forall\; \Omega: |\Omega| = K,  i: j_i\in S^{K-1} \backslash \Omega. 
\end{align}

\end{definition}
When $K=n$ or 1, $S^{K-1} \backslash \Omega=\emptyset$, it is
natural to define   $\alpha^G = 0$. 
It is easy to observe that   $\alpha^G \leq \alpha$.
%Our generalized curvature measures  how close a nondecreasing  function is from being 
%\emph{supermodular}.  
Note that the  classical \emph{total} curvature     is   
  $\alpha^{\text{\upshape total}} \coloneqq 1- \min_{i\in \groundset} \frac{\rho_i(\groundset \setminus \{i\})}{\rho_i(\emptyset)}$.   
\begin{remark}\label{ob_curvature}
For a nondecreasing function $F(\cdot)$, it holds: 
a)  $\alpha, \alpha^G \in [0, 1]$;
b)  $F(\cdot)$ is
\emph{supermodular}  iff $\alpha  = 0$; 
c)  If $F(\cdot)$ is submodular, 
then $\alpha^G \leq \alpha =   \alpha^{\text{\emph{total}}}$.
\end{remark}
So for a submodular function, our  notion of curvature is 
consistent with $\alpha^{\text{\upshape total}}$. Notably,  $\alpha^G$ usually characterizes the problem better than  $\alpha^{\text{total}}$, as will be  validated  in \cref{sec_exp}. 

%% file: guarantee_proof.tex
%!TEX root = mainfile_icml17_cameraready.tex 

\section{Approximation Guarantee}
\label{sec_main_theory}
%In this section, we present our main theoretical result in Theorem~\ref{thm_21},
%providing  approximation guarantees
%of the \algname{Greedy} algorithm for maximizing nondecreasing (possibly)
%non-submodular  functions characterized by the (generalized)  submodularity ratio and  curvature. 

We present 
  approximation guarantee
of  \algname{Greedy}  in \cref{thm_21}.
Note that both  versions 
of the  submodularity ratio and  curvature apply in the proof. For brevity, we use 
$\gamma$ and $\alpha$ to refer to any of these versions in the sequel. In \cref{sec_tightness} we prove tightness of the  approximation guarantees. All omitted proofs are given in  \cref{app_proof_guarantee}.

\begin{theorem}
\label{thm_21} 
Let $F(\cdot)$ be a non-negative nondecreasing set function  with  submodularity ratio $\gamma\in [0,1]$ and  curvature $\alpha\in [0,1]$. 
The \algname{Greedy} algorithm enjoys the following  approximation guarantee for solving problem~\eqref{eq1}: 
\begin{align}\notag
%\small 
%\hspace{-0.08cm}
%  F(S^K)\!\! \geq\! \frac{1}{\alpha}\!\! \left[\!1\!\!-\!\! \left(\! \!\frac{K-\alpha\gamma}{K}\!\!\right)^K \right]\!\! F(\opt)\!\! \geq\!\!
%\frac{1-e^{-\alpha\gamma}}{\alpha}\!F(\opt), \hspace{-0.18cm}
  F(S^K) & \geq \frac{1}{\alpha} \left[1-\left( \frac{K-\alpha\gamma}{K} \right)^K \right] F(\opt)\\ \label{eq_guarantee}
  &  \geq 
\frac{1}{\alpha}{(1-e^{-\alpha\gamma})}F(\opt),
\end{align}
where $\opt$ is  the optimal solution of~\eqref{eq1} and $S^K$ the output of the \algname{Greedy} algorithm.\footnote{For the setting that  \algname{Greedy} is allowed to pick more than $K$ 
elements, e.g., pick $K'> K$ elements, our theory can be easily 
extended to show that $F(S^{K'}) \geq   {\alpha^{-1}}(1-e^{-\alpha\gamma{K'}/{K}})F(\opt)$.}
\end{theorem}

\subsection{Interpreting \cref{thm_21}}

\begin{figure}[tbp]
%\vspace{-.7em}
\center
  \includegraphics[width=0.4\textwidth]{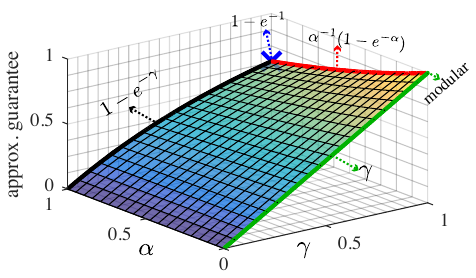}
\caption{Approximation guarantee $\frac{1}{\alpha}(1-e^{-\alpha\gamma})$. The blue cross marks the classical $(1-1/e)$-guarantee of \algname{Greedy}. The \emph{red} line illustrates the influence of the 
curvature on the guarantees for submodular functions, and the \emph{black} line illustrates the influence of $\gamma$ on the guarantees  for the worst-case curvature $\alpha=1$. The \emph{green} line is the  guarantees for $K$-cardinality constrained  \emph{supermodular} maximization.
}
\label{fig_bounds}
\vspace{-0.5em}
\end{figure}

Before proving the theorem, we want to give the reader an intuition of the results and
show how our results 
recover and extend  several  classical   guarantees for  the \algname{Greedy} algorithm. For the case $\alpha = 0$ (i.e., $F(\cdot)$ is \emph{supermodular}), the approximation guarantee is $\lim\limits_{\alpha \rightarrow 0}\frac{1}{\alpha}(1-e^{-\alpha\gamma}) = \gamma$, 
which gives the first guarantee of greedily maximizing 
a nondecreasing supermodular function with bounded $\gamma$.
When $\gamma =1$, (i.e., $F(\cdot)$ is submodular), we recover  the guarantee of ${\alpha^{-1}}(1-e^{-\alpha})$ \citep{conforti1984submodular}.
For the case $\alpha = 1$, we have a guarantee of $(1-e^{-\gamma})$ \citep{das2011submodular}. 
For the case $\alpha =1, \gamma =1$, we recover the 
classical guarantee of $(1-1/e)$ \citep{nemhauser1978analysis}. 
We plot the constant-factor approximation guarantees for different values of
$\gamma$ and $\alpha$ in \cref{fig_bounds}. 
%
%\setkeys{Gin}{width=0.25\textwidth}
One interesting 
phenomenon is that $\gamma$ and $\alpha$ play different
roles: Looking at $\gamma=0$,
 the approximation factor is always 0, independent of the value $\alpha$
takes. In contrast, for $\alpha=0$, 
the approximation guarantee is $(1-e^{-\gamma})$. This can be interpreted as the curvature
\emph{boosting} the guarantees. 

\subsection{Proof of  \cref{thm_21}}

{
The high-level proof framework is based on  \citet{conforti1984submodular} (where they derive the approximation guarantee for maximizing a nondecreasing \emph{submodular} function with bounded curvature). However, adapting the proof to non-submodular functions requires several changes detailed in    \cref{sec_related}.}

\textbf{Proof overview.}
%Given a ground set $\groundset$, 
Let us denote all problem instances of maximizing a 
non-negative nondecreasing  function $F(\cdot)$ s.t.\ $K$-cardinality constraint ($ \max_{|S|\leq K} F(S)$)
 to be $\P_{K, \alpha, \gamma}$, 
% \footnote{Without loss of generality,   in the proof we may assume $2K\leq n$, same as the  Reduction 1 in \citet{buchbinder2014submodular}.}, 
 where $F(\cdot)$
 is parametrized by submodularity ratio $\gamma$ and 
 curvature $\alpha$. 
Let $P_{\opt, S^K} \in \P_{K, \alpha, \gamma}$ denote 
those problem instances with optimal solution $\opt$ and greedy solution $S^K$. 
We 
group all  problem instances  $\P_{K, \alpha, \gamma}$ according to the set 
$\opt\cap S^K \coloneqq \{ l_1 = j_{m_1}, l_2=j_{m_2}, \ldots, l_s=j_{m_s} \}$, where
$ j_{m_1}, \ldots, j_{m_s}$ are consistent with the order of greedy selection. 
%Note that $s$ can take values from  $0, \ldots, K$. 
Let us denote  the problem instances with $\opt\cap S^K= \{ l_1, \ldots, l_s\}$ as the group $\P_{K, \alpha, \gamma}(\{l_1,\ldots, l_s\})$. 

The \emph{main} idea of the proof is to investigate the worst-case approximation
ratio of each group of the problem instances $\P_{K, \alpha, \gamma}(\{l_1,\ldots, l_s\}), \forall \{l_1,\ldots, l_s\}\subseteq S^K$. We do this by constructing LPs based on the properties of the problem instances. 
By studying the structures of these LPs, we will prove that the 
worst-case approximation ratio of all problem instances
occurs when $\opt\cap S^K = \emptyset$. Thus  the desired approximation guarantee corresponds to the worst-case  approximation ratio of $\P_{K, \alpha, \gamma}(\emptyset)$.

\textbf{The proof.}
When $\gamma=0$ or $F(\opt) = 0$, \labelcref{eq_guarantee}
holds naturally. In the following, let $\gamma\in (0, 1]$ and $F(\opt)>0$. 
First, we present \cref{lem_1},  which
will be used to construct the LPs.
\begin{lemma}[]\label{lem_1}
  For any $\Omega\subseteq \groundset$ with $|\Omega| = K$ and any $t \in \{0, \ldots, K-1\}$, let $w^t \coloneqq |S^t\cap \Omega|$. It     holds that
  \begin{flalign}\notag 
 \alpha \sum_{i:j_i\in S^t\backslash \Omega} \rho_i + \sum_{i: j_i\in S^t\cap \Omega}\rho_i + {\gamma^{-1}}(K-w^t)\rho_{t+1} \geq     F(\Omega).
  \end{flalign}
\end{lemma}
We now specify the constructing of the LPs: For any problem instance $P_{\opt, S^K} \in \P_{K, \alpha, \gamma}(\{l_1,\ldots, l_s\})$, 
we know that $F(S^K) = \sum_{i=1}^K \rho_i$ (telescoping sum). Hence, the approximation 
ratio is $\frac{F(S^K) }{F(\opt)} = \sum_i \frac{\rho_i}{F(\opt)}$, which we denote as $R(\{l_1,\ldots, l_s\}) = \sum_i \frac{\rho_i}{F(\opt)}$. 
Define $x_i \coloneqq \frac{\rho_i}{F(\opt)}, i \in [K]$. Since
$F$ is nondecreasing, $x_i\geq 0$. 
Plugging $\Omega = \opt$  into \cref{lem_1}, and 
considering  $t = 0,\ldots, K-1$, we have in total
$K$ constraints over  the variables $x_i$, which
constitute the constraints of the LP. 
So the worst-case
approximation ratio of the group $\P_{K, \alpha, \gamma}(\{l_1,\ldots, l_s\})$   is:
\begin{align}\notag 
  \underline R(\{l_1,\ldots, l_s\}) = \min \sum\nolimits_{i=1}^{K} x_i, \text{  s.t. }x_i \geq 0  \text{ and, }
\end{align}
\vspace{-1.2em}
\footnotesize 
%\scriptsize
\setcounter{MaxMatrixCols}{20}
%\begin{wrapfigure}[14]{l}[\dimexpr\columnwidth+\columnsep\relax]{32cm}
\begin{align}  
\hspace{-.3cm}
\label{bigmatrix}
& \begin{matrix}
\text{ row } (0)\\
\text{ row }$(1)$\\
\vdots\\
\text{ row }(l_1 - 1)\\
\text{ row }(l_2-1)\\
\text{ row }(q = l_r)\\
\vdots\\
\text{ row }(l_s - 1)\\
\vdots\\
\text{ row }(K-1)
\end{matrix}
\hspace{1em}
\begin{bmatrix}
\frac{K}{\gamma}\\
 \alpha &  \frac{K}{\gamma}\\
 \vdots & \vdots & \ddots\\
 \alpha & \alpha & \cdots &  {\frac{K}{\gamma}}&  &  &  & \textbf{0} \\
  \alpha & \alpha & \cdots & 1 &  \frac{K-1}{\gamma}\\
   \alpha & \alpha & \cdots & 1 & 1 & \frac{K-r}{\gamma} \\
   \vdots & \vdots &      & \vdots & \vdots & \vdots & \ddots\\
   \alpha & \alpha & \cdots & 1 & 1 & \alpha & \cdots  & \frac{K-s + 1}{\gamma}\\ 
   \vdots & \vdots & & \vdots &\vdots &\vdots && \vdots & \ddots \\
\alpha & \alpha & \cdots & 1 & 1 & \alpha & \cdots &1 & \cdots  & \frac{K-s}{\gamma}
\end{bmatrix}
\cdot 
\begin{bmatrix}
 x_1\\
 x_2\\
\vdots\\
x_{l_1}\\
x_{l_2}\\
x_{q+1}\\
\vdots\\
x_{l_s}\\
\vdots\\
x_K
\end{bmatrix}
\geq 
\begin{bmatrix}
 1 \\
 1\\
\vdots\\
1\\
1\\
1\\
\vdots\\
1\\
\vdots\\
1
\end{bmatrix}
\end{align}
%\end{wrapfigure}
\normalsize
The following Lemma
presents the key structure of the constructed  LPs, which will be used
to deduce the relation between the LPs of different
problem instance groups.
% $\P_{K, \alpha, \gamma}(\opt\cap S^K), \forall (\opt\cap S^K) \subseteq S^K$.
\begin{lemma}\label{lem_qq1}
Assume that the optimal solution of the constructed LP is $\x^*\in \R^K_+$
% i.e., $\underline R(\{l_1,\cdots, l_s\})= \sum_i x^*_i$.  
and that  $s =  |\opt\cap S^K| \geq 1$. For all $1\leq r \leq s$ it holds that
$x^*_q \leq x^*_{q+1}$, where $q = l_r$.
\end{lemma}
\vspace{-0.1cm}
\begin{proof}[Proof sketch of \cref{lem_qq1}]
 Assume by virture of creating a contradiction that $x^*_q >  x^*_{q+1}$. 
 We  can always create a new feasible solution $\y^*\in \R_+^K$ by 
  decreasing  $x^*_q$ by some  $\epsilon >0$, while  increasing all the $x^*_{q+1}$ to $x^*_K$ by some
  proper values, s.t. $\y^*$ has smaller LP objective value. 
Specifically, we define  $\y^*$  as: for $k= 1,\ldots, q-1, y^*_k := x^*_k$; $y^*_q := x^*_q - \epsilon$; for $k = q+1, \ldots, K, y^*_k := x^*_k + \epsilon_k$ where $\epsilon_k$s are defined recursively as:  $\epsilon_{q+1} = \epsilon\frac{\gamma}{K-r}$, and 
  \begin{align}\notag 
% & \epsilon_{q+1} = \epsilon\frac{\gamma}{K-r}\\\notag 
& \epsilon_{q+1+u} = \epsilon_{q+u}\frac{K-r-u + 1 - \gamma}{K-r-u}, 1\leq u \leq K-q-1.
  \end{align} 
 \begin{claim}\label{claim1}
a) The  new solution $\y^* \geq 0$; b) 
All of the  constraints in (\ref{bigmatrix}) are still feasible for  $\y^*$.
 \end{claim}
 After that the change of the LP objective is, 
 \begin{flalign}\notag
&\Delta_{LP}= - \epsilon +\epsilon_{q+1} +\epsilon_{q+2} + \ldots + \epsilon_{K}.
%=  \\\notag
%& \epsilon [ -1 +  \frac{\gamma}{K-r} +  \frac{\gamma}{K-r}\cdot \frac{K-r-\gamma}{K-r-1} +\cdots \\\notag 
%& +   \frac{\gamma}{K-r}\cdot \frac{K-r-\gamma}{K-r-1}\cdots \frac{K-r -m+2-\gamma}{K-r-m+1} ], 
 \end{flalign}
One can prove that the LP objective  decreases:
 \begin{claim}\label{claim_332}
 For all $K\geq 1$, $1\leq r\leq q <  K$, it holds that 
 $\Delta_{LP} \leq 0, \forall \gamma\in (0, 1]$.  Equality is achieved when $r =q$ and $\gamma = 1$.
 \end{claim}
 \vspace{-0.4cm}
Therefore we reach the contradiction that $\x^*$ is an optimal solution of the constructed LP. 
\end{proof}
\vspace{-0.9em}
Given \cref{lem_qq1}, we prove in the following Lemma, which states 
 that the 
worst-case approximation ratio of all problem instances
occurs when $\opt\cap S^K = \emptyset$. 
\begin{lemma}\label{lem_34}
For all  $\{l_1,\ldots, l_s \}\subseteq S^K$, it holds that $\underline R(\{l_1,\ldots, l_s\}) \geq \underline{R}(\emptyset) =  \frac{1}{\alpha} \left[1- \left(\frac{K-\alpha\gamma}{K}\right)^K\right]$.
\end{lemma}
\vspace{-0.4cm}
So the greedy solution   has objective  $F(S^K) \geq$
$\frac{1}{\alpha} \left[1- \left(\frac{K-\alpha\gamma}{K}\right)^K\right] F(\opt)$
$\geq  \frac{1}{\alpha}(1-e^{-\alpha\gamma})F(\opt)$.
\begin{wrapfigure}[11]{l}[\dimexpr\columnwidth+\columnsep\relax]{12.3cm}
\vspace{12cm}	
\end{wrapfigure}

\vspace{-0.19cm}
\subsection{Tightness Result}\label{sec_tightness}

We demonstrate that the approximation guarantee in \cref{thm_21} is
tight, i.e., for every submodularity ratio $\gamma$ and every curvature $\alpha$, there exist set functions that achieve the bound exactly. 
%In the following we describe the construction of these functions.

Assume  the ground set $\groundset$ contains the elements in $S:=\{j_1,\ldots, j_K \}$ and the elements in $\Omega:=\{\omega_1,\ldots, \omega_K \}$ ($S\cap \Omega = \emptyset$) and  $n - 2K$ dummy elements. 
The objective function we are going to construct will
not depend on these dummy elements, i.e., the objective value of a set does not change if dummy elements are removed from or added to that set.
Consequently, the dummy elements  will not affect the submodularity ratio and the
curvature.
%W.l.o.g,  we assume that any subset $T\subseteq \groundset$ can be represented as $T = S'\cup \Omega'$, where $S'\subseteq S, \Omega'\subseteq \Omega$. 
For the constants $\alpha\in [0,1], \gamma \in (0,1]$, we define the objective function as, 
%$\forall T\subseteq \groundset$, 
\normalsize
 {
\begin{equation}\label{tight_fn} 
\small 
F(T) := 
% \left[1-\frac{\alpha\gamma}{K}f(|\Omega'|)\right]\sum_{j_i\in S'}\rho_i +\frac{f(|\Omega'|)}{K} \\\label{tight_fn}
 \frac{f(|\Omega\cap T|)}{K}\big(1-\alpha\gamma\sum_{i:j_i\in S\cap T}\xi_i \big) + \sum_{i:j_i\in S\cap T}\xi_i, 
\end{equation}}
where $\xi_i \coloneqq \frac{1}{K} \left(\frac{K-\gamma\alpha}{K} \right)^{i-1}, i \in [K]$; $f(x) = \frac{\gamma^{-1} -1}{K-1}x^2 + \frac{K-\gamma^{-1}}{K-1}x$. Note that  $f(x)$ is  convex nondecreasing   over $[0, K]$, and that $f(0)=0, f(1) = 1, f(K) = K/\gamma$. It is clear that $F(\emptyset) = 0$ and $F(\cdot)$ is monotone
nondecreasing.  The following lemma shows that it is 
generally non-submodular and non-supermodular. 
\begin{lemma}\label{claim45}
For the objective in \labelcref{tight_fn}: a) When $\alpha =0$, it is supermodular; b) When $\gamma=1$, it is submodular; c) $F(T)$  has submodularity ratio $\gamma$ and 
curvature $\alpha$.
\end{lemma}
%So in general $F(T)$ is non-submodular and non-supermodular. 
Considering the problem of $\max_{|T|\leq K} F(T)$, 
we claim that the \algname{Greedy} algorithm may output $S$. 
%with  value $F(S) = \frac{1}{\alpha\gamma}\left[1- \left(\frac{K-\alpha\gamma}{K}\right)^K\right]$. 
This can be proved
by  induction. 
One can see that $\rho_{j_1}(\emptyset) =\xi_1=\rho_{\omega_1}(\emptyset)$, 
so  \algname{Greedy}  can choose $j_1$ in the first step. 
Assume in step $t-1$  \algname{Greedy}   has chosen $S^{t-1} = \{j_1, \ldots, j_{t-1}\}$, one can  verify that the marginal gains coincide, i.e., 
%\begin{flalign}\notag 
$\rho_{j_t}(S^{t-1}) = \xi_t =   \rho_{\omega_t}(S^{t-1})$.
%= \frac{f(1)}{K}\left(1- \alpha\gamma \sum_{i=1}^{t-1} \rho_{t-1}\right) =  \frac{1}{K} \left(\frac{K-\gamma\alpha}{K} \right)^{t-1}
%\end{flalign}
However,  the optimal solution is actually $\Omega$ with function value as $F(\Omega) = \frac{1}{\gamma}$. 
So the approximation ratio is $\frac{F(S)}{F(\Omega) } = \frac{1}{\alpha}\left[1- \left(\frac{K-\alpha\gamma}{K}\right)^K\right] $, which 
matches our approximation guarantee in \cref{thm_21}.

%% file: app.tex
%!TEX root = mainfile_icml17_cameraready.tex 
\section{Applications}

We consider several important real-world applications and their corresponding objective functions. We show that
the submodularity ratio and the curvature of these functions can be bounded and, hence, the approximation guarantees
from our theoretical results are applicable. All the omitted
proofs are provided in \cref{sec_app_proof_app}.

\subsection{Bayesian A-optimality in Experimental Design}

In Bayesian experimental design \citep{chaloner1995bayesian}, the goal is to
select a set of experiments to perform s.t.\ some statistical
criterion is optimized, e.g., the variance of certain
parameter estimates is minimized.
\citet{krause2008near} investigated several criteria for this purpose, amongst others the 
Bayesian  A-optimality criterion.
This criterion is used to 
maximally reduce the variance in the posterior 
distribution over the parameters.
In general, the criterion is \emph{not} submodular as shown in \citet[Section 8.4]{krause2008near}. % provide a counter example

Formally, assume there are $n$ experimental stimuli $\{\x_1,\ldots, \x_n \}$, each $\x_i\in \R^d$,
which constitute the data matrix $\bmX\in \R^{d\times n}$.
 Let us arrange a set $S\subseteq \groundset$ 
of stimuli as a matrix $\bmX_S \coloneqq [\x_{v_1}, \ldots, \x_{v_s}]	 \in \R^{d\times |S|}$. Let $\bmtheta \in 
\R^d$ be the parameter vector in the linear model $\y_S = \bmX_S^{\trans}\bmtheta + \w$, where $\w$ 
is the Gaussian noise with zero mean and variance $\sigma^2$, i.e., $\w \sim \N(0, \sigma^2 \bmI)$, and $\y_S$ is the vector of dependent variables.  
Suppose the prior takes the form of an isotropic Gaussian, i.e., 
$\bmtheta \sim \N(0, \bmLambda^{-1}), \bmLambda = \beta^2\bmI$. 
Then,
\begin{align}\notag 
 \begin{bmatrix}
 \y_S\\
 \bmtheta
 \end{bmatrix}
 \sim \N(0, \bmSigma), \bmSigma = 
 \begin{bmatrix}
 \sigma^2 \bmI + \bmX_S^\trans\bmLambda^{-1}\bmX_S &  \bmX_S^\trans\bmLambda^{-1}\\
 \bmLambda^{-1}\bmX_S & \bmLambda^{-1}
 \end{bmatrix}.
\end{align}
This implies that $\bmSigma_{\bmtheta | \y_S} = (\bmLambda + \sigma^{-2}\bmX_S \bmX_S^{\trans} )^{-1}$.
%\begin{flalign}\notag 
%&	\bmSigma_{\bmtheta | S} \coloneqq \bmSigma_{\bmtheta | \y_S} = (\bmLambda + \sigma^{-2}\bmX_S \bmX_S^{\trans} )^{-1}.
%%\\\notag 
%%	&  = \bmLambda^{-1} - \bmLambda^{-1}\bmX_S(\sigma^2\bmI + \bmX_S^\trans\bmLambda^{-1}\bmX_S)^{-1}\bmX_S^\trans \bmLambda^{-1}.
%\end{flalign}
The A-optimality objective is defined as,
\begin{flalign}\label{eq_bayesian_a_op}
	F_A(S) &\coloneqq \tr{\bmSigma_{\bmtheta}}-\tr{\bmSigma_{\bmtheta|\y_S}} \\\notag 
& 	= \tr{\bmLambda^{-1}}-\tr{(\bmLambda + \sigma^{-2}\bmX_S \bmX_S^{\trans} )^{-1}}.
\end{flalign}
The following \namecref{lemma_a_opt} gives bounds on the submodularity ratio and curvature of \labelcref{eq_bayesian_a_op}.
\begin{proposition}\label{lemma_a_opt}
Assume normalized stimuli, i.e., $\|\x_i\| = 1, \forall i\in \groundset$.
  Let the spectral norm of $\bmX$ be $\|\bmX\|$.\footnote{By 
    Weyl's inequality, a naive upper bound is $\|\bmX\| \leq \sqrt{n}$.} 
% W.l.o.g.\ assume that $\beta = \sigma = 1$. 
 Then, 
a) The  objective in  \labelcref{eq_bayesian_a_op} is  monotone nondecreasing. b)  Its  submodularity ratio $\gamma$ can be lower bounded by $\frac{\beta^2}{\|\bmX\|^2(\beta^2+\sigma^{-2}\|\bmX\|^2)}$,
and its  curvature $\alpha$ can be upper bounded by $1 - \frac{\beta^2}{\|\bmX\|^2(\beta^2+\sigma^{-2}\|\bmX\|^2)}$.
\end{proposition}

\subsection{The Determinantal Function}

The determinantal function of a square submatrix is widely used in many areas, e.g., in determinantal 
point processes \citep{kulesza2012determinantal} and active set selection for sparse Gaussian processes. 
%A DPP on a set of items $\groundset$ is a probability 
%measure on the power set $2^{\groundset}$. 
%Let $\bmL$  be a positive semidefinite matrix, for every $S\subseteq \groundset$ we have
%$P(S) \propto \de{\bmL_S}$. 
%The MAP inference of a DPP is to maximize $P(S)$, or equivalently maximizing $\de{\bmL_S}$, which is non-monotone in general. 
%\sebastian{@Andrew: As the objective for DPPs is non-montone, should we even talk about them? This is kind of misleading.}
%
%We consider monotone nondecreasing determinantal functions in this work, which, for example, 
Monotone nondecreasing determinantal functions 
appear in the second problem. Assume $\bmSigma$ is the covariance matrix parameterized by a positive
definite kernel. In the Informative Vector
Machine \citep{lawrence2003fast}, the information gain of a subset of points $S\subseteq \groundset$ is $\frac{1}{2}\log F(S)$, where
\begin{align}\label{eq_det}
  F(S) \coloneqq \de{\bmI + \sigma^{-2}\bmSigma_S},
\end{align}
where $\sigma$ is the noise variance in the Gaussian process model, $\bmSigma_S$ is the square submatrix
 with both its rows and columns indexed by $S$. 
Although 
$\log F(S)$ is  submodular,
%\citep{Krause05near-optimalnonmyopic}, 
$F(S)$ is in general not submodular.
The approximation guarantee of  \algname{Greedy}  for maximizing $\log F(S)$ does not
translate to a guarantee for maximizing $F(S)$. The following \namecref{prop_ratio_determinant} characterizes   \eqref{eq_det}.
% Note
%that  $\bmI + \sigma^{-2}\bmSigma_S = (\bmI + \sigma^{-2}\bmSigma)_S$
%is a symmetric positive definite matrix. 
\begin{proposition}\label{prop_ratio_determinant}
a)  $F(S)$ in \labelcref{eq_det} is  supermodular, its curvature is 0; b) 
Let the eigenvalues of $\bmA := \bmI + \sigma^{-2}\bmSigma$
be $\lambda_1\geq \cdots \geq \lambda_n > 1$. 
The greedy submodularity ratio of $F(S)$ can be lower bounded by
$\frac{K(\lambda_n - 1)}{(\prod_{j=1}^{K}\lambda_j) -1}$.
\end{proposition}

\subsection{LPs with Combinatorial Constraints}
\label{app_lp}

LPs with combinatorial constraints appear frequently in practice.
Consider the following example: Suppose that $\groundset$ is the set of all products a company can produce.
Given budget constraints on the raw materials needed, companies consider the LP $ \max_{\x\in \P} \dtp{\d}{\x}$, where $\d$ is the vector of profits for the individual products and where $\P$ is
a polytope representing the continuous constraints.
The above LP can be used to assess the profit maximizing production plan. 
Usually the company needs to consider \emph{combinatorial} constraints
as well. For instance,  the company has at most $K$ production lines, thus they have to select a subset of
 $K$ products to produce.
%  $S: |S|\leq K$. All the products $\groundset$ may also
%be partitioned into different categories, the company may want to
%choose some products from each category so as to maintain the
%diversity of products. This can be modeled by a partition matroid constraint, where each category of products constitutes one %block
%in the partition matroid. 
%
Often  this kind of problems can be formalized as  $ \max_{\x\in \P, \spt{\x}\in \I} \dtp{\d}{\x}$, where $\I$ is the independent set of the  combinatorial
structure. Hence, a natural  auxiliary set function is,
\begin{flalign}\label{lp_objective}
F(S) := \max\nolimits_{\spt{\x}\subseteq S,  \; \x\in \P} \dtp{\d}{\x}, \; \forall S\subseteq \groundset.
\end{flalign}
Let $\P = \{\x\in \R^n \;|\; 0\leq \x\leq \bar \u, \bmA\x \leq \b, \bar \u\in \R_+^n, \bmA\in \R_+^{m\times n}, \b\in \R_+^m \}$. 
In general $F(S)$ in  \labelcref{lp_objective} is non-submodular as illustrated by
 two examples in  \cref{app_subsec_lp}.
Upper bounding the curvature   is equivalent to lower bounding $\frac{F(S \cup \Omega) -  F(S\setminus \{i\} \cup \Omega)}{F(S) - F(S\setminus \{i\})}$,
which  can be 0 in the worst case. However, the submodularity ratio
can be lower bounded by a non-zero scalar.
\begin{proposition}[]\label{lemma_lp_ratio}
a) $F(S)$ in \labelcref{lp_objective} is a normalized nondecreasing  set function. 
b) With  regular non-degenerancy assumptions (details in   \cref{app_subsub_lp}), 
   its  submodularity ratio can be  lower bounded by $\gamma_0 >0$. 
\end{proposition}

\subsection{More Applications}
Many  real-world applications
 can benefit from the theory in
this work, for instance: subset selection using 
the $R^2$ objective, sparse modeling and the budget allocation
problem with combinatorial constraints. 
Details on these applications are deferred   to
\cref{app_more_apps}.

%% file: exp.tex
%!TEX root = main_ratio_curvature_icml_format.tex
%\vspace{-0.4cm}
\setkeys{Gin}{width=0.23\textwidth}
 \begin{figure}[htbp]

% \center 
 \subfloat[Objective values/OPT \label{fig_boston_exp_design1}]{
 \includegraphics[]{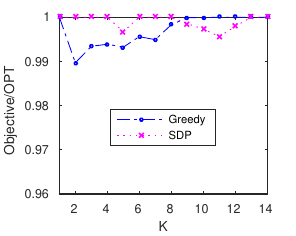}
\hspace{-0.4cm}}
 \subfloat[Parameters \label{fig_boston_exp_design2}]{
 \includegraphics[]{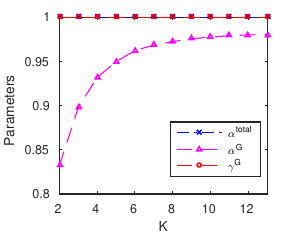}}
 \caption{Results  on the Boston Housing data.}
 \label{fig_boston_exp_design}
 \vspace{-0.2cm}
 \end{figure}

\section{Experimental Results}
\label{sec_exp}
We empirically validated  approximation guarantees
characterized by the submodularity 
ratio and the curvature for several applications. 
Since it is too time consuming to calculate the full versions of  $\alpha$ and $\gamma$ using exhaustive search, we only calculated the \emph{greedy}
versions ($\alpha^G, \gamma^G$). All  averaged results are from   20 repeated experiments. Source code is available at \url{https://github.com/bianan/non-submodular-max}.\footnote{All experiments were
implemented using Matlab.
We used the SDP solver provided by  CVX (Version 2.1). } More results are put in \cref{more_exps}.

\subsection{Bayesian Experimental Design}

We considered the Bayesian A-optimality objective for both synthetic  and 
real-world data. In all experiments, we normalized the data points to have unit $\ell_2$-norm. 

 \setkeys{Gin}{width=0.23\textwidth}
 \begin{figure}[htbp]
% \vspace{-0.5cm}
 \center 
 \addtocounter{subfigure}{-2}
 \subfloat{
 \includegraphics[]{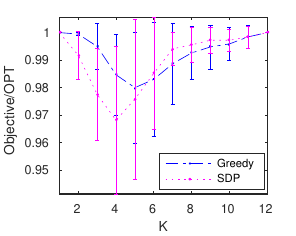}
 }
 \subfloat{
 \includegraphics[]{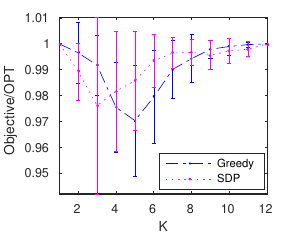}}\\
 \vspace{-0.69cm}
 \subfloat[Correlation: 0.2 \label{fig_syn_a_bar_1}]{
 \includegraphics[]{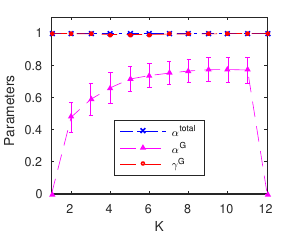}}
 \subfloat[Correlation: 0.6 \label{fig_syn_a_bar_2}]{
 \includegraphics[]{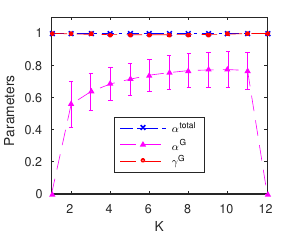}}
 \caption{Results for A-optimality on synthetic data.}
 \label{fig_syn_a_bar}
 \end{figure}
\textbf{Real-world results:} 
We used the
 Boston Housing Data. 
The dataset\footnote{\url{https://archive.ics.uci.edu/ml/datasets/Housing}} has $14$ features (e.g., crime rate, property tax rates, etc.) and  $516$ samples. 
 To be able to quickly calculate the parameters and optimal solution by 
 exhaustive search,
% we used the first $d= 6$ features and 
 the first $n = 14$ samples were used.
 As a baseline, we used an SDP-based algorithm (abbreviated as \algname{SDP},  details are available  in \cref{sdp_a_opt}).  
 Results are shown in \cref{fig_boston_exp_design} for varying values of $K$.
 In \cref{fig_boston_exp_design1} we can observe that both  \algname{Greedy} and {\algname{SDP}} compute near-optimal solutions.
 From \Cref{fig_boston_exp_design2} we can see that the greedy
 submodularity ratio $\gamma^G$ is close to 1, and that the greedy
 curvature $\alpha^G$ is less than 1, while the classical
 curvature $\alpha^{\text{total}}$ is always 1 (the worst-case value). This implies that the classical total curvature $\alpha^{\text{total}}$
 characterizes the considered maximization problems less accurate than  the greedy curvature. 
% \vspace{-0.5cm}

\textbf{Synthetic results:} 
We generated 
random observations from a multivariate Gaussian distribution with different 
correlations.
To be able to assess
the ground truth,  we used  $n=12$ samples with $d=6$ features.
   \cref{fig_syn_a_bar} shows the results with correlation $0.2$ (first column) and $0.6$ (second column), respectively: The first row shows the average objective
   values over the optimal value with error bars, and the second row  shows the parameters. One can
   observe that \algname{Greedy} always obtains near-optimal solutions and
   that these solutions are roughly comparable with  those obtained by the  \algname{SDP}. The classical curvature $\alpha^{\text{total}}$  is always close to 1, while $\alpha^G$
   take smaller values,  and $\gamma^G$ takes values close to 1,   thus characterize the  performance of 
   \algname{Greedy} better. 

  \setkeys{Gin}{width=0.23\textwidth}
  \begin{figure}[tbp]
%   \vspace{-0.4cm}
  \center 
  \subfloat[$n=112, d=40, \text{corr.}= .5$ \label{subfig_det_1}]{
  \includegraphics[]{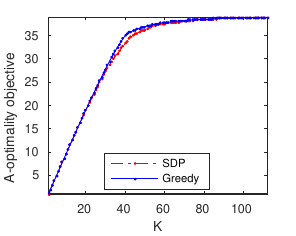}}
  \subfloat[$n=48, d=24, \text{corr.}:.99$ \label{subfig_det_2}]{
  \includegraphics[]{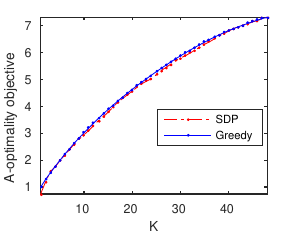}}\\
  \caption{A-optimality on medium-scale problems}
  \label{fig_exp_large}
  \end{figure}
 \textbf{Medium-scale synthetic experiments:} To
 compare the runtime of \algname{SDP} and \algname{Greedy},
we considered  \emph{medium-scale} datasets (we cannot report results on larger datasets because of the huge computational demands of the  \algname{SDP}). 
  \Cref{fig_exp_large} shows the objective value 
  achieved by \algname{Greedy}
     and  \algname{SDP} for different numbers of features $d$ and numbers of samples $n$, as well as the correlations. We can observe  that  \algname{Greedy} 
     computes solutions that are on par or superior to those of \algname{SDP}.  
 In \cref{tab_timing} we summarize the runtime  of \algname{Greedy}
   and  \algname{SDP} for different values of $d$ and $n$, for correlation $0.5$. Furthermore, we show  
   the ratio of runtimes of the two algorithms. We can observe that  \algname{Greedy}
   is usually two \emph{orders} of magnitude faster than \algname{SDP}. 
  \begin{table}[htbp]
  \footnotesize 
  	\centering
  	  	\caption{Runtime in seconds of  \algname{Greedy}
  	  		and  \algname{SDP}. The last row shows the ratio of
  	  		runtimes of \algname{SDP} and \algname{Greedy}.}
%  \hspace{-0.85cm}
  	  	\label{tab_timing}  
\begin{tabular}{| l | l | l | l | l | l  | l | l | l | l  }
  		\hline 
  		 & $d$: 60 & $d$: 40 & $d$: 64 & $d$: 100 & $d$: 120   \\
  		   & $n$: 80& $n$: 112 & $n$: 128 & $n$: 200 & $n$: 250\\ 
  		\hline 
  		\hline 
  		\algname{Greedy}    &   0.278 &  0.360 &   0.765 & 4.666 & 10.56  \\
  		\hline 
  		\algname{SDP}        &  95.2& 115.2 &   205.4 & 1741.2  & 3883.5  \\
  		\hline 
  		$\frac{ \algname{SDP}}{\algname{Greedy} }$& 
  		341.7& 319.9 &  268.7 & 373.2 & 367.7  \\
  		\hline   
  	\end{tabular}
  \end{table}

 \subsection{LPs with Combinatorial Constraints}

\setkeys{Gin}{width=0.23\textwidth}
\begin{figure}[htbp]
\vspace{-0.2cm}
   \center 
       \addtocounter{subfigure}{-2}
   \subfloat{
   \includegraphics[]{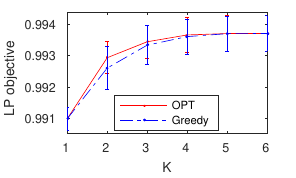}}
   \subfloat{
   \includegraphics[]{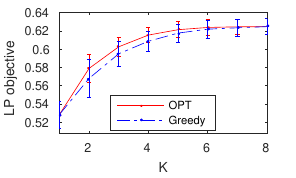}}\\
   \vspace{-.698cm}
   \subfloat[$n=6, m=20$ \label{subfig_lp_1}]{
   \includegraphics[]{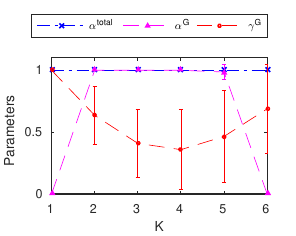}}
    \subfloat[$n=8,m=30$ \label{subfig_lp_2}]{
      \includegraphics[]{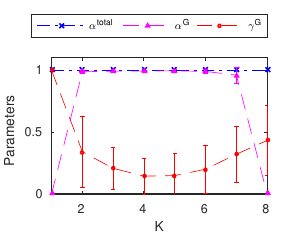}}
   \caption{Results  for LPs with
   $K$-cardinality constraints.}
   \label{fig_syn_lp}
 \vspace{-0.4cm}
   \end{figure}

 We generated synthetic LPs as follows: 
 Firstly,   we generated the matrix $\bmA\in \R^{m\times n}_+, A_{ij}\in [0, 1]$ by
 drawing all entries independently from a uniform distribution on $[0,1]$.
  We set 
   $\b =  \d= \mathbf{1}$, and  set $\bar \u$ as $\mathbf{1}$.      
% The function value and parameters averaged over 
% 20 repeated experiments are shown in .
%  Since $\alpha^{\text{classical}}$ is almost always 1, so
%  we omit plotting it. 
The first row of \cref{fig_syn_lp}  plots the optimal LP objective (calculated
 using exhaustive search) and the LP objective returned by
  \algname{Greedy}. The second row 
 shows the  curvature and submodularity ratio.
 The first column (\cref{subfig_lp_1}) presents the results
 for $n=6, m=20$, while the second column (\cref{subfig_lp_2})
 presents that  for $n=8,m=30$. 
Note the greedy submodularity ratio
 takes values between $\sim 0.15$ and $1$, and that the curvature is close 
 to the worst-case value of $1$. These observations are consistent
 with the theory in \cref{app_lp}.
% 
% the approximation bounds give a reasonable prediction
% of the performance of \algname{Greedy} algorithm. 

 \subsection{Determinantal Functions Maximization}
  
\begin{figure}[htbp]
  \center 
  \subfloat{
  \includegraphics[]{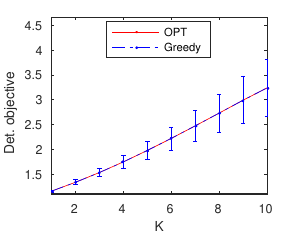}}
  \subfloat{
  \includegraphics[]{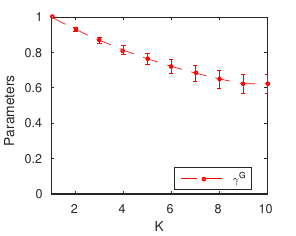}}\\
  \caption{Synthetic result. Left: objective value, right: $\gamma^G$}
  \label{fig_det}
  \end{figure}

  We experimented with synthetic and real-world data: 
  For synthetic data, we generated random covariance matrices $\bmSigma\in \R^{n\times n}$ with uniformly
  distributed eigenvalues in $[0, 1]$. We set $n = 10, \sigma = 2$. 
  %
%  \Cref{fig_det} plots the results on 20 repeated experiments. 
  In \cref{fig_det} (left) we plot the optimal determinantal objective
  value and the value achieved by  \algname{Greedy}. 
  \Cref{fig_det} (right) traces the greedy submodularity ratio $\gamma^G$. 
  Since the determinantal objective is  supermodular,  so the approximation guarantee equals to $\gamma^G$. 
  We can see that $\gamma^G$ can reasonably predict the performance 
   of \algname{Greedy}.
  \setkeys{Gin}{width=0.23\textwidth}
 \vspace{-0.3cm}
 \vspace{-0.4cm}
 
\begin{figure}[htbp]
   \center 
\vspace{-0.3cm}
   \subfloat{
   \includegraphics[]{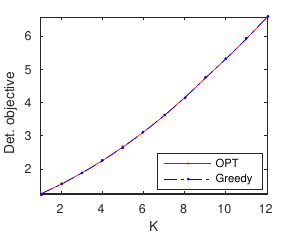}}
   \subfloat{
   \includegraphics[]{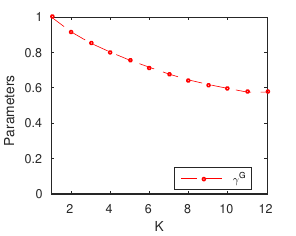}}\\
   \caption{CIFAR-10 result. Left: objective value, right: $\gamma^G$}
   \label{fig_det_cifar}
 \vspace{-0.2cm}
\end{figure}

 For real-world data, we considered an active set
 selection task on the  CIFAR-10\footnote{\url{https://www.cs.toronto.edu/~kriz/cifar.html}} dataset. The first $n = 12$ images in the test
 set were used to calculate the covariance matrix with
 an squared exponential kernel ($k(\x_i, \x_j)=\exp(-\|\x_i - \x_j\|^2/h^2)$, $h$ was set to be 1). The results in  \Cref{fig_det_cifar} shows   similar 
 results as with the synthetic data. 
   \setkeys{Gin}{width=0.2\textwidth}

%% file: related.tex
%!TEX root = mainfile_icml17_cameraready.tex 

\section{Related Work}
\label{sec_related}

In this section we briefly discuss  related work on
 various notions of non-submodularity and
 the optimization of   non-submodular functions (Further  details   in \cref{sec_related_work}).

\textbf{Relation to \citet{conforti1984submodular} in deriving approximation guarantees.}   {
In proving \cref{thm_21}  we use the similar proof framework (i.e., utilizing  LP formulations to analyze the worst-case approximation ratios of different groups of problem instances) as that in \citet{conforti1984submodular},  where they derive guarantees for maximizing submodular functions. 
However, since we are proving guarantees for  non-submodular functions,  the specific techniques on how to manipulate these LPs are different. Specifically, 1) The building block to construct  LPs (\cref{lem_1}) is different;
2) The  technique to prove the structure of the LPs (which corresponds to \cref{lem_qq1}) is significantly different for a submodular function and a  non-submodular function, and \cref{lem_qq1} is 
the  key to investigate the worst-case approximation ratios of different groups of problem instances. 3) The specific way to prove \cref{lem_34} is also different since the constraints of the  LPs are different for submodular and non-submodular functions. }

\textbf{Submodularity ratio and curvature.} Curvature
% \citep{conforti1984submodular,vondrak2010submodularity,iyer2013curvature} 
 is typically  defined for submodular
functions.  \citet{sviridenko2015optimal}  present a notion
of curvature for monotone non-submodular functions. 
  \cref{app_classical_defs} provides  
details of that  notion and relates it to our definition.
% Additionally, we prove in \cref{append_remark} that our combination of  (generalized) curvature and submodularity ratio  is more expressive than  that of \citet{sviridenko2015optimal} in characterizing the problem~\eqref{eq1}.  
\citet{yoshida2016maximizing} prove an improved 
approximation ratio for knapsack-constrained 
maximization of  submodular functions with bounded
curvature. 
 {Submodularity ratio} \citep{das2011submodular}
 is  a quantity characterizing how {close} a function is to being submodular.

\textbf{Approximate submodularity.}
\citet{krause2008near}
define \emph{approximately submodular}  functions with parameter  $\epsilon\geq 0$ 
as those functions $F$ that satisfy an approximate 
diminishing returns property, i.e.,  
$\forall A\subseteq B \subseteq \groundset\setminus v$ it holds that
  $\rho_v(A) \geq \rho_v(B) -\epsilon$.
 \algname{Greedy}   yields a solution with objective 
$F(S^K) \geq (1-e^{-1})F(\Omega^*) - K\epsilon$,
for maximizing a monotone $F$ s.t.\ a $K$-cardinality 
constraint.
\citet{du2008analysis}  study  the greedy maximization of non-submodular 
potential functions with \emph{restricted submodularity} and \emph{shifted submodularity}. Restricted submodularity refers to functions which are submodular only over some
collection of subsets of $\groundset$,  and shifted submodularity can be viewed as a special case
of the approximate diminishing returns as defined above. %\cref{eq_approx_dr}. 
Recently, \citet{NIPS2016_6236} study  \emph{$\epsilon$-approximately submodular}
functions, which arised from their research on  ``noisy" submodular functions. 
A function $F(\cdot)$
is $\epsilon$-approximately submodular if there exists a submodular function
$G$ s.t.\  $(1-\epsilon)G(S)\leq F(S)\leq (1+\epsilon)G(S)$,  $\forall S\subseteq \groundset$.
% \algname{Greedy}  achieves an approximation guarantee of
%$O(1-1/e - \delta)$, where $0 \leq \delta < 1$, for sufficiently small $\epsilon$ for monotone submodular functions~\citep[Theorem 5]{NIPS2016_6236}.
%
%\citet{altschuler2016greedy} analyzed the approximation 
%guarantee of the \algname{Greedy} algorithm on the 
%column subset selection problem. 

\textbf{Weak submodularity.}
\citet{borodin2014weakly} study  \emph{weakly submodular} functions, i.e.,
 montone, nomalized  functions $F(\cdot)$ s.t.
for any  $S$,  $T$, it holds $|T|F(S) + |S|F(T)\geq |S\cap T|F(S\cup T) + |S\cup T|F(S\cap T)$. 
For a function $F(\cdot)$, 
we show in \cref{counter_ex} that  the following two facts do not imply each other: i) 
$F(\cdot)$  is weakly  submodular; ii) The submodularity ratio of
$F(\cdot)$ is strictly larger than 0, and its curvature is strictly smaller than  1.

\textbf{Other notions of non-submodularity.}  \citet{feige2013welfare} introduce the \emph{supermodular degree} as a complexity measure
for set functions. They show that a greedy
algorithm for the welfare maximization problem
enjoys an approximation guarantee increasing linearly
with the supermodular degree. 
\citet{NIPS2016_6384} use  the \emph{submodularity index} to characterize the performance 
of the \algname{RandomGreedy} algorithm \citep{buchbinder2014submodular} for maximizing a non-monotone function.
%Further discussions are in   \cref{sec_related_work}. 

\textbf{Optimization of    non-submodular functions.}
The submodular-supermodular procedure has
been proposed to minimize the difference of two
submodular functions \citep{narasimhan2005submodular,iyer2012algorithms}. 
\citet{jegelka2011submodularity} present the problem 
of minimizing ``cooperative cuts", which are non-submodular in general, and propose efficient algorithms for optimization. 
\citet{kawahara2015approximate} analyze unconstrained minimization of the sum of a submodular function and a tree-structured supermodular function. 
\citet{bai2016algorithms} investigate the minimization of the 
ratio of two submodular functions, which can be solved with bounded approximation factor.

%\if 0
%\paragraph{Constrained submodular minimization.}
%For size-constrained submodular minization, there is the 
%$\sqrt{\frac{n}{\log n}}$ hardness for $\min\{F(S): |S|\geq K \}$ \citep{svitkina2011submodular}. 
%
%\fi 

%% file: disc.tex
%!TEX root = main_ratio_curvature_icml_format.tex
\vspace{-0.1cm}
\section{Conclusion}
\vspace{-0.1cm}
We  analyzed the
guarantees for greedy maximization of  
non-submodular nondecreasing set functions. 
By combining   the (generalized) curvature $\alpha$ and submodularity
ratio $\gamma$ for generic  set functions, we prove the 
\emph{first} tight  approximation bounds  in terms of these definitions for greedily maximizing nondecreasing set functions. These approximation bounds significantly enlarge the  domain where   \algname{Greedy}  has guarantees. 
Furthermore, we theoretically bounded  the parameters $\alpha$ and $\gamma$ for several non-trivial
applications, and validate our theory in various experiments.

%For future work, we plan to prove approximation
%guarantee for non-submodular function maximization
%subject to more general constraints, e.g.,  matroid and
%knapsack constraints. The other direction
%is to bound the curvature of the auxiliary set function
%for feature selection with strongly convex loss function.  

%\begin{itemize}
%
%
%\item  Add the approximate update oracle \citep{goundan2007revisiting}
%
%
%\item Weakly submodular functions by \cite{borodin2014weakly}, actually quite different. 
%
%%\item Discrete greedy (coordinate ascent) may be able to deal with SFMax
%%with both combinatorial and convex constraints. 
%%
%%The algorithm of  \cite{DBLP:journals/mor/Wolsey82} and \cite{soma2014optimal}
%%can be treated as in this manner. 
%%
%%Disadvantage:  difficult to deal with complex constraints. 
%
%
%\item   bound the curvature for general strongly convex
%and smooth function. 

%\end{itemize}

%% file: appendix.tex
%!TEX root = mainfile_icml17_cameraready.tex

\newpage 
\onecolumn

\begin{center}
\LARGE
\textbf{Appendix}
\end{center}

\section{Organization of the Appendix}

\Cref{app_proof_guarantee} presents the proofs for our approximation guarantees and its tightness for the 
\algname{Greedy} algorithm.

\Cref{app_classical_defs} provides details on existing  notions of curvature  and submodularity ratio, and relates it
to the notions in this paper. 

\Cref{sec_app_proof_app} presents detailed proofs for 
bounding the  submodularity ratio and curvature
for various applications.  

\Cref{sdp_a_opt} gives details on the classical SDP formulation
of the  Bayesian A-optimality objective. 

\Cref{sec_related_work} provides proofs omitted in \cref{sec_related}. 

\Cref{app_more_apps} provides information on more applications, including
sparse modeling  with strongly convex loss functions, subset 
selection using the $R^2$ objective and optimal  budget allocation
with combinatorial constraints. 

\Cref{more_exps} provides  experimental results on subset selection
with the $R^2$ objective and additional results on experimental design. 

\section{Proofs for  Approximation Guarantee and Tightness Result (\cref{sec_defs} and \cref{sec_main_theory})}
\label{app_proof_guarantee}

\subsection{Proof of Remarks in \cref{sec_defs}}

\begin{proof}[Proofs of \cref{obs_ratio}]
\mbox{ }

a)  Because $F$ is nondecreasing,  and $\gamma, \gamma^G$ are defined as the largest scalars, 
$\gamma, \gamma^G \geq 0$. 
At the same time, both
$\gamma$ and $\gamma^G$ can be at most 1 because the conditions in \cref{def:gen-submod-ratio} also have to hold for the case
that 
$\Omega\backslash S$ ($\Omega\backslash S^t$, respectively) is a singleton.

b) 
%We know that $\gamma$ cannot
% be greater than 1 (consider $\Omega\setminus S$ 
% being a singleton).    
 $``\Rightarrow"$: 
 
 Let  $\Omega\setminus S = \{\omega_1,\ldots, \omega_k \}, k\geq 1$. 
 % From submodularity, we know that $F(\{\omega_i\} \cup S) + F(\{\omega_j\} \cup S ) \geq F(\{\omega_i, \omega_j \}\cup S) + F(S)$, so we have $\sum_{i=1}^{k}\rho_{\omega_i}(S) \geq \rho_{\Omega}(S) $, so $\gamma$ can take the largest  value  1. 
 Submodularity implies $\sum_{i=1}^{k}\rho_{\omega_i}(S) \geq \rho_{\Omega}(S) $. Hence, $\gamma$ can take the largest  value  1. 
 
  $``\Leftarrow"$: 
 
 $\gamma =1$ implies that (setting $\Omega\setminus S=\{\omega_i , \omega_j \}$), for all $\omega_i, \omega_j \in \groundset \setminus S$, it holds that $F(\{\omega_i\} \cup S) + F(\{\omega_j\} \cup S ) \geq F(\{\omega_i, \omega_j \}\cup S) + F(S)$, which is an equivalent way to define submodularity \citep[Proposition 2.3]{bach2013learning}. 

\end{proof}

\begin{proof}[Proof of \cref{ob_curvature}]\mbox{ }

 a) ``If $F(\cdot)$ is 
nondecreasing, then $\alpha, \alpha^G \in [0, 1]$";

When $\Omega = \emptyset$,  $\alpha$ 
is  at least 0. From the definition, $\alpha^G \geq 0$. 
Since $F$ is nondecreasing,
$ \rho_{i}(S\setminus \{i\} \cup \Omega)\geq 0$ (respectively, ${\rho_{j_i}(S^{i-1}\cup \Omega)}\geq 0$), 
 and we defined
$\alpha, \alpha^G$ to be the smallest scalar, it must hold that $\alpha, \alpha^G \leq 1$.

b) ``For a nondecreasing function $F(\cdot)$,  $F(\cdot)$ is
supermodular  iff $\alpha  = 0$ ";

 $``\Rightarrow"$: 
 
 If $F$ is supermodular, it always holds that $ \rho_{i}(S\setminus \{i\} \cup \Omega) \geq {\rho_{i}(S\setminus \{i\})}$, 
 combined with the fact that 
  $\alpha$ is at least 0, we know that $\alpha$ must be 0.

  $``\Leftarrow"$:

One can observe that $\alpha=0$ is equivalent to
$-F(\cdot)$ satisfying the diminishing returns property, which is
equivalent to $F(\cdot)$ being supermodular.

c)  ``If $F(\cdot)$ is nondecreasing submodular, 
then $\alpha^G \leq \alpha =  \alpha^{\text{total}}$."

Since it always holds that $\alpha^G \leq   \alpha$, we only need to prove that  $ \alpha = \alpha^{\text{total}}$. 
Wlog., assume $\rho_{i}(S\setminus \{i\})>0$. Then,
\begin{align}\notag 
1-\alpha & = \min_{ \Omega, S\subseteq \groundset, i\in S \backslash \Omega}\frac{\rho_{i}(S\setminus \{i\} \cup \Omega)}{\rho_{i}(S\setminus \{i\})}\\\notag 
& =  \min_{S\subseteq \groundset, i\in S}\frac{\rho_{i}(\groundset\setminus \{i\})}{\rho_{i}(S\setminus \{i\})} \quad \text{ (diminishing returns, and taking $\Omega=\groundset\setminus\{i \}$)}\\\notag
& =  \min_{i\in \groundset}\frac{\rho_{i}(\groundset\setminus \{i\})}{\rho_{i}(\emptyset)} \quad \text{ (diminishing returns, and taking $S = \{i\}$)}\\\notag  \
& = 1 -\alpha^{\text{total}}.
\end{align}
So it holds that  $\alpha^G \leq \alpha = \alpha^{\text{total}}$.
\end{proof}

\subsection{Proof of \cref{lem_1}}

\begin{proof}[Proof of \cref{lem_1}]

The proof needs the definitions of generalized  curvature, submodularity ratio,  and the  selection rule of the \algname{Greedy} algorithm. 

%Remember that we use the shorthand  . 
Firstly, observe,
\begin{flalign}
F(\Omega\cup S^t)  &= F(\Omega) + \sum_{i: j_i\in S^t} \rho_{j_i} (\Omega\cup S^{i-1}) \nonumber\\
  &= F(\Omega) + \sum_{i: j_i\in S^t\backslash \Omega} \rho_{j_i} (\Omega\cup S^{i-1}) + \underbrace{\sum_{i: j_i\in S^t \cap \Omega} \rho_{j_i} (\Omega\cup S^{i-1})}_{\textnormal{$=0$ because $j_i \in \Omega$}} \nonumber\\
  &= F(\Omega) + \sum_{i: j_i\in S^t\backslash \Omega} \rho_{j_i} (\Omega\cup S^{i-1}). \label{obser}
\end{flalign}

%Firstly let us prove the following identity, 
%\begin{flalign}\label{obser}
%F(\Omega\cup S^t)  = F(\Omega) + \sum_{i: j_i\in S^t\backslash \Omega} \rho_{j_i} (\Omega\cup S^{i-1}).
%\end{flalign}
%We know that $S^t = \{j_1, \cdots, j_t \}$, wlog., let 
%$S^t\setminus \Omega = \{j_1,\cdots, j_a, j_b, \cdots, j_t \}, a\leq b-1$, and $S^t \cap \Omega = \{j_{a+1}, \cdots,  j_{b-1} \}$.  By telescoping we know that, 
%\begin{align}\notag 
%& \sum_{i =1}^a \rho_{j_i} (\Omega\cup S^{i-1})  = F(\Omega\cup S^a) - F(\Omega) \quad \textnormal{and}\\\notag 
% & \sum_{i =b}^t \rho_{j_i} (\Omega\cup S^{i-1})  = F(\Omega\cup S^t) - F(\Omega\cup S^{b-1}). 
%\end{align}
%Because $S^t \cap \Omega = \{j_{a+1}, \cdots,  j_{b-1} \}$,
%it holds that  $F(\Omega\cup S^{b-1})  = F(\Omega\cup S^a) $, so $ \sum_{i: j_i\in S^t\backslash \Omega} \rho_{j_i} (\Omega\cup S^{i-1}) =\sum_{i =1}^a \rho_{j_i} (\Omega\cup S^{i-1}) +  \sum_{i =b}^t \rho_{j_i} (\Omega\cup S^{i-1}) = F(\Omega\cup S^t)  - F(\Omega) $, which proves \labelcref{obser}. 

From the definition of the submodularity ratio, 
\begin{flalign}\label{eq_ratio_used}
F(\Omega\cup S^t) \leq F(S^t)  + \frac{1}{\gamma} \sum_{\omega\in \Omega\backslash S^t} \rho_{\omega}(S^t).
\end{flalign}
From the definition of curvature (for the greedy curvature, since 
it holds for $S^{K-1}$, it must also hold for $S^t\subseteq S^{K-1}$), we have, 
\begin{flalign}\label{eq_cur_used}
 \sum_{i: j_i\in S^t\backslash \Omega} \rho_{j_i} (\Omega\cup S^{i-1})  \geq (1-\alpha)  \sum_{i: j_i\in S^t\backslash \Omega} \rho_{j_i} (S^{i-1}).
\end{flalign}
Combining \labelcref{eq_ratio_used,obser,eq_cur_used}, and remember that we use the shorthand $\rho_t \coloneqq \rho_{j_t}(S^{t-1})$, it reads, 
\begin{flalign}\notag 
 F(\Omega) &= F(\Omega\cup S^t) - \sum_{i: j_i\in S^t\backslash \Omega} \rho_{j_i} (\Omega\cup S^{i-1}) \\\notag 
& \leq \alpha \sum_{i:j_i\in S^t\backslash \Omega} \rho_i +F(S^t) -   \sum_{i: j_i\in S^t\backslash \Omega}\rho_i +  \frac{1}{\gamma} \sum_{\omega\in \Omega\backslash S^t} \rho_{\omega}(S^t) \\\notag
&= \alpha \sum_{i:j_i\in S^t\backslash \Omega} \rho_i + \sum_{i: j_i\in S^t\cap \Omega}\rho_i +  \frac{1}{\gamma} \sum_{\omega\in \Omega\backslash S^t} \rho_{\omega}(S^t)\\\notag
& \leq  \alpha \sum_{i:j_i\in S^t\backslash \Omega} \rho_i + \sum_{i: j_i\in S^t\cap \Omega}\rho_i + \gamma^{-1}(K-w^t)\rho_{t+1},
\end{flalign}
where the last inequality is because of the selection rule of the \algname{Greedy} algorithm ($\rho_{\omega}(S^t) \leq \rho_{t+1}, \forall \omega$). 
\end{proof}
%Combining \labelcref{eq_ratio_used,obser,eq_cur_used}, we have,
%\begin{flalign}\notag 
% F(\Omega) 
%& \leq \alpha \sum_{i:j_i\in S^t\backslash \Omega} \rho_i +F(S^t) -   \sum_{i: j_i\in S^t\backslash \Omega}\rho_i +  \frac{1}{\gamma} \sum_{\omega\in \Omega\backslash S^t} \rho_{\omega}(S^t)\\\notag
%&= \alpha \sum_{i:j_i\in S^t\backslash \Omega} \rho_i + \sum_{i: j_i\in S^t\cap \Omega}\rho_i +  \frac{1}{\gamma} \sum_{\omega\in \Omega\backslash S^t} \rho_{\omega}(S^t)\\\notag
%& \leq  \alpha \sum_{i:j_i\in S^t\backslash \Omega} \rho_i + \sum_{i: j_i\in S^t\cap \Omega}\rho_i + \gamma^{-1}(K-w^t)\rho_{t+1},
%\end{flalign}
%where the last inequality is because of the selection rule of the \algname{Greedy} algorithm ($\rho_{\omega}(S^t) \leq \rho_{t+1}, \forall \omega$). 
%\end{proof}

\subsection{Proof of \cref{claim1}}

Since the proof heavily relies on the structure 
of the constructed LPs, we restate it here:
The worst-case
approximation ratio of the group $\P_{K, \alpha, \gamma}(\{l_1,..., l_s\})$   is
\begin{align}\notag 
  \underline R(\{l_1,..., l_s\}) = \min \sum\nolimits_{i=1}^{K} x_i, \text{    s.t.   }x_i \geq 0, i\in [K]  \text{ and }
\end{align}

\setcounter{MaxMatrixCols}{20}
\begin{align}  \notag 
& \begin{matrix}
\text{ row } (0)\\
\text{ row }$(1)$\\
\vdots\\
\text{ row }(l_1 - 1)\\
\text{ row }(l_2-1)\\
\text{ row }(q = l_r)\\
\vdots\\
\text{ row }(l_s - 1)\\
\vdots\\
\text{ row }(K-1)
\end{matrix}
\hspace{1em}
\begin{bmatrix}
K/\gamma\\
 \alpha &  K/\gamma\\
 \vdots & \vdots & \ddots\\
 \alpha & \alpha & \cdots &  {K/\gamma}&  &  &  & \textbf{0} \\
  \alpha & \alpha & \cdots & 1 &  (K-1)/\gamma\\
   \alpha & \alpha & \cdots & 1 & 1 & \frac{K-r}{\gamma} \\
   \vdots & \vdots &      & \vdots & \vdots & \vdots & \ddots\\
   \alpha & \alpha & \cdots & 1 & 1 & \alpha & \cdots  & \frac{K-s + 1}{\gamma}\\ 
   \vdots & \vdots & & \vdots &\vdots &\vdots && \vdots & \ddots \\
\alpha & \alpha & \cdots & 1 & 1 & \alpha & \cdots &1 & \cdots  & \frac{K-s}{\gamma}
\end{bmatrix}
\cdot 
\begin{bmatrix}
 x_1\\
 x_2\\
\vdots\\
x_{l_1}\\
x_{l_2}\\
x_{q+1}\\
\vdots\\
x_{l_s}\\
\vdots\\
x_K
\end{bmatrix}
\geq 
\begin{bmatrix}
 1 \\
 1\\
\vdots\\
1\\
1\\
1\\
\vdots\\
1\\
\vdots\\
1
\end{bmatrix}
\labelcref{bigmatrix}
\end{align}

For notational simplicity, w.l.o.g., assume that 
$j_i = i, i\in [K]$. Let the row index in (\ref{bigmatrix}) 
start from 0. 

\begin{proof}[Proof of \cref{claim1}]\mbox{ }

Let  $\epsilon = \frac{(K-r + 1-\gamma)x^*_q - (K -r)x^*_{q+1}}{K-r +1}  \geq  \frac{K-r}{K-r +1} (x^*_{q} - x^*_{q+1})>0$.

\textbf{a)}  It is easy to see that $\y^* \geq 0$ since 
the only decreased entry is the  $q^{\text{th}}$ entry, and one can easily see  that 
$y^*_q = x^*_{q}  - \epsilon \geq 0$. 

\textbf{b)} ``All of the  constraints in (\ref{bigmatrix}) are still feasible for  $\y^*$.''

\textbf{(i)} For the rows $0$ to $(q-2)$ in  (\ref{bigmatrix}), there is 
no change, so they are still feasible. 

\textbf{(ii)}
For the $(q-1)^{\text{th}}$
 and $q^{\text{th}}$ rows in (\ref{bigmatrix}), they  are 
 \begin{flalign}\label{eq_loose}
 \alpha x^*_1 + \cdots + \alpha (\text{or } 1)x^*_{q-1}  & + \frac{K-r+1}{\gamma} x^*_q         \geq 1\\\label{eq_loose_1}
  \alpha x^*_1 + \cdots + \alpha (\text{or } 1)x^*_{q-1}  & +x^*_q   + \frac{K-r}{\gamma}x^*_{q+1}    \geq 1
 \end{flalign}
 
For  \labelcref{eq_loose}, after plugging  $\y^*$ into
its L.H.S., we get $ \alpha y^*_1 + \cdots + \alpha (\text{or } 1)y^*_{q-1}  + \frac{K-r+1}{\gamma} y^*_q $, subtract from 
which the L.H.S. of \labelcref{eq_loose_1}, we get 
 \begin{flalign}\notag 
& \left[ \alpha y^*_1 + \cdots + \alpha (\text{or } 1)y^*_{q-1}  + \frac{K-r+1}{\gamma} y^*_q\right] - \left[ \alpha x^*_1 + \cdots + \alpha (\text{or } 1)x^*_{q-1}   +x^*_q   + \frac{K-r}{\gamma}x^*_{q+1}\right]\\ \notag 
& = \frac{K-r+1}{\gamma} (x^*_q - \epsilon) - x^*_q   - \frac{K-r}{\gamma}x^*_{q+1} \\\notag 
& = 0, 
 \end{flalign}
so $ \alpha y^*_1 + \cdots + \alpha (\text{or } 1)y^*_{q-1}  + \frac{K-r+1}{\gamma} y^*_q \geq 1$ and $\y^*$ is feasible 
for \labelcref{eq_loose}.
 
  After increasing $x^*_{q+1}$ by 
  $\epsilon_{q+1} = \epsilon\frac{\gamma}{K-r}$, the $q^{\text{th}}$ row in (\ref{bigmatrix}) is feasible since the change in its L.H.S. is $-\epsilon + \epsilon = 0$.

\textbf{(iii)}
For the rows $q$ to $(K-1)$ in  (\ref{bigmatrix}), let 
 us prove  by \textbf{induction}. 
 
 For the base case, consider the $(q+1)^{\text{th}}$ row in (\ref{bigmatrix}), it can be either, 
 \begin{flalign}\notag 
   \alpha x^*_1 + \cdots + \alpha (\text{or } 1)x^*_{q-1}  & +x^*_q   + x^*_{q+1}  +  \frac{K-r-1}{\gamma}x^*_{q+2}    \geq 1\\\notag 
   \text{or}\\\notag
    \alpha x^*_1 + \cdots + \alpha (\text{or } 1)x^*_{q-1}  & +x^*_q   + \alpha x^*_{q+1}  +  \frac{K-r}{\gamma}x^*_{q+2}    \geq 1
 \end{flalign}
 It can be easily verified that the $(q+1)^{\text{th}}$ row in (\ref{bigmatrix}) is still feasible 
 in both the above two situations. 
 Let us use $\Delta_{q+u}$ to denote the change of L.H.S. of the $(q+u)^{\text{th}}$ row
 after applying  the changes. 
 
 For the inductive step, assume that the claim holds for $u = u'$, i.e., the $(q+u')^{\text{th}}$ row in (\ref{bigmatrix}) is  feasible or $\Delta_{q+u'} \geq 0$.  The $(q+u')^{\text{th}}$ row
 is,
 \begin{flalign}\notag 
\texttt{(...same as $(q+u' +1)^{\text{th}}$ row) } + \frac{K-r-v}{\gamma}x^*_{q+u'+1} \geq 1 
 \end{flalign}
 where $0\leq v\leq u'$ is some integer dependent on the structure of (\ref{bigmatrix}), but not affect the final analysis. 
 Then the $(q+u'+1)^{\text{th}}$ row
 can be either,
  \begin{flalign}\notag 
&\texttt{(... same as $(q+u')^{\text{th}}$ row) } + 
  x^*_{q+u'+1} + \frac{K-r-v-1}{\gamma}x^*_{q+u'+2}   \geq 1  \texttt{ (case 1)}\\\notag 
    \text{or}\\\notag 
& \texttt{(... same as  $(q+u')^{\text{th}}$ row) } + \alpha x^*_{q+u'+1} + \frac{K-r-v}{\gamma}x^*_{q+u'+2}    \geq 1 \texttt{ (case 2)}
  \end{flalign}
In \texttt{ (case 1)},  the L.H.S. of  $(q+u'+1)^{\text{th}}$ row  minus  the L.H.S. of  $(q+u')^{\text{th}}$ row  is $\frac{K-r-v-1}{\gamma}x^*_{q+u'+2}  - \frac{K-r-v-\gamma}{\gamma}x^*_{q+u'+1}$, so
\begin{flalign}\notag 
 \Delta_{q+u'+1} - \Delta_{q+u'} 
& = \frac{K-r-v-1}{\gamma}\epsilon_{q+u'+2}  - \frac{K-r-v-\gamma}{\gamma}\epsilon_{q+u'+1}\\\notag
& = \left[\frac{K-r-v-1}{\gamma}\frac{K-r-u' -\gamma}{K-r-u'-1}  - \frac{K-r-v-\gamma}{\gamma}\right]\epsilon_{q+u'+1}\\\notag
& = \left[({K-r-v-1})\frac{K-r-u' -\gamma}{K-r-u'-1}  - ({K-r-v-\gamma})\right]\frac{\epsilon_{q+u'+1}}{\gamma}\\\notag
& \geq  \left[({K-r-v-1})\frac{K-r-v -\gamma}{K-r-v-1}  - ({K-r-v-\gamma})\right]\frac{\epsilon_{q+u'+1}}{\gamma}  {\qquad    \texttt{(since $0\leq v\leq u'$)}}\\\notag 
 & = 0.
\end{flalign}
so the $(q+u'+1)^{\text{th}}$ row is still feasible. 

In \texttt{ (case 2)}, the L.H.S. of  $(q+u'+1)^{\text{th}}$ row  minus  the L.H.S. of  $(q+u')^{\text{th}}$ row  is $\frac{K-r-v}{\gamma}x^*_{q+u'+2}  - (\frac{K-r-v}{\gamma} - \alpha)x^*_{q+u'+1}$, so
 \begin{flalign} \notag 
 \Delta_{q+u'+1} - \Delta_{q+u'}
  & =\frac{K-r-v}{\gamma}\epsilon_{q+u'+2}  - (\frac{K-r-v}{\gamma} - \alpha)\epsilon_{q+u'+1}\\\notag
 & \geq \frac{K-r-v}{\gamma}(\epsilon_{q+u'+2} - \epsilon_{q+u'+1}) \quad \texttt{(since $\alpha\geq 0$)}\\\notag
 & \geq 0.  \quad   \texttt{(since $\epsilon_{q+u'+2} \geq  \epsilon_{q+u'+1}$)}
 \end{flalign}
 so the $(q+u'+1)^{\text{th}}$ row is feasible. Thus we finish proving  \cref{claim1}. 
 \end{proof}
 
 \subsection{Proof of \cref{claim_332}}
 
  \begin{proof}[Proof of \cref{claim_332}]

 The change of the LP objective is
  \begin{flalign}\notag
 \Delta_{LP}&= - \epsilon +\epsilon_{q+1} +\epsilon_{q+2} + \cdots + \epsilon_{K}
  \\\notag
 &= \epsilon \left[ -1 +  \frac{\gamma}{K-r} +  \frac{\gamma}{K-r}\cdot \frac{K-r-\gamma}{K-r-1} +\cdots
%  \\\notag 
% & \;\;\;\;
 +   \frac{\gamma}{K-r}\cdot \frac{K-r-\gamma}{K-r-1}\cdots \frac{K-r -m+2-\gamma}{K-r-m+1} \right], 
  \end{flalign}
  where inside the bracket there are $m = K-q$ items except for the $``-1"$.   
   For notational simplicity, let the sum inside the bracket to be, 
 \begin{flalign} 
    & h_r(\gamma) :=  -1 +  \frac{\gamma}{K-r} +  \frac{\gamma}{K-r}\cdot \frac{K-r-\gamma}{K-r-1}
%    \\\notag
%    &  
    +\cdots +   \frac{\gamma}{K-r}\cdot \frac{K-r-\gamma}{K-r-1}\cdots \frac{K-r -m+2-\gamma}{K-r-m+1}.
   \end{flalign}
  
  First of all, since $K-r \geq K-q = m$,
  we have that
  \begin{flalign}\label{subeq_39}
  \small 
 & h_r(\gamma) \leq h_{r=q}(\gamma) = -1 +   \frac{\gamma}{m} +  \frac{\gamma}{m}\cdot \frac{m-\gamma}{m-1} +\cdots +  \frac{\gamma}{m}\cdot \frac{m-\gamma}{m-1} \cdots \frac{3-\gamma}{2}\cdot\frac{2-\gamma}{1}.
  \end{flalign}
  Let us merge the items in \labelcref{subeq_39} from left to right one by one, 
  \begin{flalign}\notag
 % \small 
    h_{r=q}(\gamma)
    = &-1 + \frac{\gamma}{m} +  \frac{\gamma}{m}\cdot \frac{m-\gamma}{m-1} +\cdots  +  \frac{\gamma}{m}\cdot \frac{m-\gamma}{m-1} \cdots \frac{3-\gamma}{2}\cdot\frac{2-\gamma}{1}\\\notag 
  = & -\frac{m-\gamma}{m} +  \frac{\gamma}{m}\cdot \frac{m-\gamma}{m-1} +\cdots  +  \frac{\gamma}{m}\cdot \frac{m-\gamma}{m-1} \cdots \frac{3-\gamma}{2}\cdot\frac{2-\gamma}{1}\\\notag
  =& -\frac{m-\gamma}{m} \frac{m-1-\gamma}{m-1}  +\cdots  +  \frac{\gamma}{m}\cdot \frac{m-\gamma}{m-1} \cdots \frac{3-\gamma}{2}\cdot\frac{2-\gamma}{1}\\\notag
 & \cdots\\\notag
  = &- \frac{(m-\gamma)(m-\gamma-1)\cdots (2-\gamma)(1-\gamma)}{m(m-1)\cdots 2\cdot 1}\\\notag
   & \overset{\text{setting $\gamma$ to be 1}}{\leq}  0
  \end{flalign}
  Then  $h_r(\gamma) \leq 0, \forall \gamma\in (0, 1]$. And it is easy to see that the equality holds if $r =q$ and $\gamma = 1$. 
  
  So we have that 
  $\Delta_{LP} =\epsilon h_r(\gamma) \leq 0$, where the equality is  achieved  at ``boundary" situation ($r =q$ and $\gamma = 1$).
  \end{proof}
  
\subsection{Proof of \cref{lem_34}}
  
\begin{proof}[Proof of \cref{lem_34}]\mbox{ }

For notational simplicity, wlog., assume that 
$j_i = i, i\in [K]$.

\textbf{a) }
Firstly let us prove that $\underline R(\{l_1,..., l_s\}) \geq \underline{R}(\emptyset)$. 

The high-level idea is to change the structure of the constraint matrix in the LP associated with $\{l_1,..., l_s \}$, such that in each change, the optimal LP objective value $\underline{R}$  never increases. 
  	
To better explain the proof, let us state the \emph{setup} first of all. 
 Let us call the elements inside the set $\opt\cap S^K =\{l_1=j_{m_1}, l_2=j_{m_2},..., l_s=j_{m_s} \}$ the 
 ``joint elements", which means that they are joint elements in 
 $\opt$ and $S^K$. 
 Similarly, 
   the elements outside of $\opt\cap S^K$ are
 called the ``disjoint" elements.  For the joint elements, two elements $l_i, l_j$ being ``adjacent" means 
 that $l_i+ 1 =l_j$. Mapping to the constraint matrix in  (\ref{bigmatrix}), it means that the corresponding 
 columns (column ($l_i$) and column ($l_j$)) are adjacent with each other.  So we also call the corresponding 
 columns in the constraint matrix as ``joint columns". 
  	
  	We prove part \textbf{a)} of \cref{lem_34} by two steps:  In the first step, we try to make all of the joint elements inside $\{l_1, l_2,..., l_s \}$ to be adjacent with each other; In the second step, we  get rid of the joint columns in the constraint matrix from left to right,  one by one. Specifically, 
  	\paragraph{Step 1.}  
  	Assume that some elements inside $\{l_1, l_2,..., l_s \}$ are 
  	not adjacent, like the example  in (\ref{bigmatrix}),
  	where $l_2$ and $l_3$ are not adjacent. Suppose that
  	$l_r$ and $l_{r+1}$ are not adjacent, which means $l_r +1< l_{r+1}$. Denote
  	$p=l_r$ for notational simplicity. Let us use $\bmA$ to represent the constraint matrix
  	in the constructed LP associated with $\{l_1, l_2,..., l_{r-1}, \textcolor{red}{l_r}, l_{r+1}, ...,  l_s \}$,
  	let $\bmA'$ represent the constraint matrix associated with $\{l_1, l_2,..., l_{r-1}, \textcolor{red}{l_r +1}, l_{r+1}, ...,  l_s \}$. Notice that $l_r+1$ is  a 
  	disjoint element for $\bmA$, but a joint element for $\bmA'$.   Furthermore $\bmA$ and $\bmA'$ only differ by columns 
  	$p$ and $p+1 = l_r +1$. 
  	Assume that $\x^*\in \R_+^K$ is the optimal solution of the constructed
  	LP with $\bmA$ as its constraint matrix. 
  	From \cref{lem_qq1}, it must hold that $x^*_p\leq x^*_{p+1}$. Combining with the fact that $\bmA\x^*\geq 1$, 
  	one can easily  verify that $\bmA'\x^* \geq 1$, which implies that,
  	\begin{flalign}\notag 
  	& \underline{R}(\{l_1, l_2,..., l_{r-1}, \textcolor{red}{l_r}, l_{r+1}, ...,  l_s \})\\
  	& \geq \underline{R}(\{l_1, l_2,..., l_{r-1}, \textcolor{red}{l_r +1}, l_{r+1}, ...,  l_s \}).
  	\end{flalign} 
  	The change from $\{l_1, l_2,..., l_{r-1}, \textcolor{red}{l_r}, l_{r+1}, ...,  l_s \}$ to  $\{l_1, l_2,..., l_{r-1}, \textcolor{red}{l_r +1}, l_{r+1}, ...,  l_s \}$ is essentially to swap the roles of  one originally  disjoint element $\textcolor{red}{l_r +1}$ and  the originally joint  element $\textcolor{red}{l_r}$.
  	Repeatedly applying this operation for all $1\leq r \leq s-1$ such that $l_r + 1< l_{r+1}$,  we can get that, 
  	\begin{flalign}\notag  
  	&\underline{R}(\{l_1, l_2,..., l_{r-1}, {l_r}, l_{r+1}, ...,  l_s \})\\
  	& \geq \underline{R}(\{l_s -s+1, l_s -s +2, ..., l_s -1,  l_s \}).
  	\end{flalign} 
  	Now the $s$ joint  elements inside $\{l_s -s+1, l_s -s +2, ..., l_s -1,  l_s \}$
  	are adjacent with each other. 
  	
  	\paragraph{Step 2.}
  	Let $\bmB$ be the constraint matrix associated with $\{l_s -s+1, l_s -s +2, ..., l_s -1,  l_s \}$, and  $\bmB'$ be the constraint matrix associated with $\{l_s -s +2, ..., l_s -1,  l_s \}$. Note that  $\bmB$ and $\bmB'$ differ in the columns from $l_s -s+1$ to the end. Suppose the vector $\x^*$ is the optimal solution of the constructed LP with $\bmB$ as 
  	the constraint matrix. 
%  	$\underline{R}(\{l_s -s+1, l_s -s +2, \cdots, l_s -1,  l_s \})$.
  	%
  	 According to \cref{lem_qq1} we know that $x^*_{l_s -s+1} \leq x^*_{l_s -s+2} \leq \cdots \leq x^*_{l_s } \leq x^*_{l_s +1}$. 
  	So one  can easily verify that it must hold that $\bmB'\x^* \geq 1$, which implies 
  	\begin{flalign}\notag 
  	&\underline{R}(\{l_s -s+1, l_s -s +2, ..., l_s -1,  l_s \})\\
  	&  \geq \underline{R}(\{l_s -s +2, ..., l_s -1,  l_s \}).
  	\end{flalign}
  	Apply this process repeatedly  $s$ times, one can reach that $\underline{R}(\{l_s -s+1, l_s -s +2, ..., l_s -1,  l_s \}) \geq \underline{R}(\emptyset)$.
  	
  	Combining step 1 and step 2, we prove part \textbf{a)} of \cref{lem_34}.
  	
  	\textbf{b) }
  	Then let us prove that $\underline{R}(\emptyset) = \frac{1}{\alpha} \left[1- \left(\frac{K-\alpha\gamma}{K}\right)^K\right]$. 
  	
  	The constructed LP  associated with $\underline{R}(\emptyset)$ is, 
  		\begin{flalign}\notag 
  		\underline R(\emptyset) = \min \sum_{i=1}^{K} x_i
  		\end{flalign}
  		subject to the constraints that, 
  		$$x_i \geq 0, \forall i = 1,... , K$$
  		and 
  		\setcounter{MaxMatrixCols}{20}
  		\begin{equation}\label{bigmatrix2} 
  		\begin{bmatrix}
  		\frac{K}{\gamma}\\
  		\alpha &  \frac{K}{\gamma}\\
  		\vdots & \vdots & \ddots\\
  		\alpha & \alpha & \cdots &  \frac{K}{\gamma}&  &  &  & \textbf{0} \\
  		\alpha & \alpha & \cdots & \alpha &  \frac{K}{\gamma}\\
  		\alpha & \alpha & \cdots & \alpha & \alpha & \frac{K}{\gamma}  \\
  		\vdots & \vdots &      & \vdots & \vdots & \vdots & \ddots\\
  		\alpha & \alpha & \cdots & \alpha & \alpha & \alpha & \cdots  & \frac{K}{\gamma}\\ 
  		\vdots & \vdots & & \vdots &\vdots &\vdots && \vdots & \ddots \\
  		\alpha & \alpha & \cdots & \alpha & \alpha & \alpha & \cdots &\alpha & \cdots  & \frac{K}{\gamma}
  		\end{bmatrix}
  		\cdot 
  		\begin{bmatrix}
  		x_1\\
  		x_2\\
  		\vdots\\
  		x_{a}\\
  		x_{b}\\
  		x_{c}\\
  		\vdots\\
  		x_{d}\\
  		\vdots\\
  		x_K
  		\end{bmatrix}
  		\geq 
  		\begin{bmatrix}
  		1 \\
  		1\\
  		\vdots\\
  		1\\
  		1\\
  		1\\
  		\vdots\\
  		1\\
  		\vdots\\
  		1
  		\end{bmatrix}
  		\end{equation}
  		One can observe that the vector $\y\in \R_+^K$ such that  $y_i = \frac{\gamma}{K}\left( \frac{K-\gamma\alpha}{K}\right)^{i-1}, i = 1,..., K$ satisfies all the constraints
  		and every row in (\ref{bigmatrix2}) is tight, hence $\y$ is the optimal solution.
  		So $$\underline R(\emptyset) = \sum_{i=1}^{K} y_i =  \frac{1}{\alpha} \left[1- \left(\frac{K-\alpha\gamma}{K}\right)^K\right].$$
  \end{proof}

  \subsection{Proof  for  the Tightness Result}
  
  \begin{proof}[Proof of \cref{claim45}]
  \mbox{ }
  
  \textbf{a)} ``When $\alpha =0$, $F(\cdot)$ is supermodular"; 
  
  It is easy to see that $\xi_i=1/K, i\in [K]$. Since $f(\cdot)$
  is convex, it can be easily verified that $F(\cdot)$ is supermodular.
  
  \textbf{b)} ``When $\gamma=1$, $F(\cdot)$ is submodular"; 
  
  Now $f(x) = x$. Assume there are $T_1\subseteq T_2 \subseteq \groundset, t\in \groundset\setminus T_2$. Let $T_1 = S_1'\cup \Omega_1', T_2 = S_2'\cup \Omega_2'$,
  where $S_1',  S_2' \subseteq S, \Omega_1', \Omega_2' \subseteq \Omega$. 
  It  holds that  $S_1'\subseteq S_2', \Omega_1'\subseteq \Omega_2'$. Now there are two cases: 
  
  1)  $t=j_i\in S$. Then, 
  \begin{align}\notag
  \rho_{j_i}(T_1) = \left[ 1- \frac{\alpha\gamma}{K}f(|\Omega_1'|) \right]\xi_i,   \qquad
  \rho_{j_i}(T_2) = \left[ 1- \frac{\alpha\gamma}{K}f(|\Omega_2'|) \right]\xi_i
  \end{align}
  Because $f(\cdot)$ is nondecreasing, so it holds $\rho_{j_i}(T_1)  \geq \rho_{j_i}(T_2)$.
  
  2)  $t= \omega_i\in \Omega$.  It reads,
  \begin{align}\notag 
  \rho_{\omega_i}(T_1) = \frac{1}{K}\left[ 1- {\alpha\gamma}\sum_{j_i\in S_1'}\xi_i  \right], \qquad  
   \rho_{\omega_i}(T_2) = \frac{1}{K}\left[ 1- {\alpha\gamma}\sum_{j_i\in S_2'}\xi_i  \right]
  \end{align}
  Because $S_1'\subseteq S_2'$, so $\rho_{\omega_i}(T_1)\geq \rho_{\omega_i}(T_2)$.
  
  The above two situations prove the submodularity of $F(T)$ when $\gamma=1$.
  
  \textbf{c)} ``$F(T)$  has submodularity ratio  $\gamma$ and 
  curvature  $\alpha$".
  
   Let us assume $T = A\cup B$ and $T' = A'\cup B'$ are two \emph{disjoint} sets ($T\cap T' = \emptyset$), where $A$ and $A'$ are subsets of $S$ while 
  $B$ and $B'$ are subsets of $\Omega$. It is easy 
  to see that $A\cap A' = \emptyset, B\cap B' = \emptyset$.

  First of all, for the \textbf{submodularity ratio},  
  assume without loss of generality\footnote{If $\rho_{T'}(T)  =0$, from monotonicity of $F(\cdot)$, it must hold $\sum_{i\in T'}\rho_i{(T)} = 0$, this case is not of interest in \cref{def:gen-submod-ratio}.} that $\rho_{T'}(T)  >0$, so the 
  submodularity ratio is $\gamma = \min_{T, T'}\frac{\sum_{i\in T'}\rho_i{(T)} }{\rho_{T'}(T) }$.
  
  One can see that, 
  \begin{flalign}\notag 
  \rho_{T'}(T) & = F(T'\cup T) - F(T) \\\notag
  &= \frac{f(|B\cup B'|) - f(|B|)}{K}(1-\alpha\gamma \sum_{j_i\in A}\xi_i) +
   \left[1-\frac{\alpha\gamma}{K}f(|B\cup B'|)  \right]\sum_{j_i\in A'}\xi_i
  \end{flalign}
  and 
  \begin{flalign}\notag
  \sum_{i\in T'}\rho_i{(T)} &= \sum_{\omega_i\in B'}\rho_{\omega_i}(T)+ \sum_{j_i\in A'}\rho_{j_i}(T) \\\notag
  & = |B'|\frac{f(|B|+1)-f(|B|)}{K}\left(1-\alpha\gamma \sum_{j_i\in A}\xi_i \right)+ 
   \left[1 - \frac{\alpha\gamma}{K}f(|B|)\right]\sum_{j_i\in A'}\xi_i.
  \end{flalign}
%  According to the definition 
%    of submodularity ratio, we have $|T'|\leq K, |T|\leq K-1$. 
%  
%  Since $\gamma$ is always less than or equal to $1$, so we can assume w.l.o.g. that
%  $\frac{\sum_{i\in T'}\rho_i{(T)}}{\rho_{T'}(T) }\leq 1$. 
  
Because $f(|B|)\leq f(|B\cup B'|)$, so one has $\left[1-\frac{\alpha\gamma}{K}f(|B\cup B'|)  \right]\sum_{j_i\in A'}\xi_i \leq \left[1 - \frac{\alpha\gamma}{K}f(|B|)\right]\sum_{j_i\in A'}\xi_i$,  equality holds when $B'=\emptyset$ or $A'=\emptyset$.  Therefore,
  \begin{flalign}\notag 
  \frac{\sum_{i\in T'}\rho_i{(T)}}{\rho_{T'}(T) } & =   \frac{|B'|\frac{f(|B|+1)-f(|B|)}{K}\left(1-\alpha\gamma \sum_{j_i\in A}\xi_i \right) +  \left[1 - \frac{\alpha\gamma}{K}f(|B|)\right]\sum_{j_i\in A'}\xi_i}{ \frac{f(|B\cup B'|) - f(|B|)}{K}(1-\alpha\gamma \sum_{j_i\in A}\xi_i) +
     \left[1-\frac{\alpha\gamma}{K}f(|B\cup B'|)  \right]\sum_{j_i\in A'}\xi_i}\\\notag 
  &  \geq \frac{|B'|\frac{f(|B|+1)-f(|B|)}{K}\left(1-\alpha\gamma \sum_{j_i\in A}\xi_i \right) +  \left[1 - \frac{\alpha\gamma}{K}f(|B|)\right]\sum_{j_i\in A'}\xi_i}{ \frac{f(|B\cup B'|) - f(|B|)}{K}(1-\alpha\gamma \sum_{j_i\in A}\xi_i) +
       \left[1-\frac{\alpha\gamma}{K}f(|B|)  \right]\sum_{j_i\in A'}\xi_i}\\\label{eq34}
  & \geq  \frac{|B'|(f(|B|+1)-f(|B|))}{f(|B\cup B'|)-f(|B|)},
  \end{flalign}
  where \labelcref{eq34} comes from the fact: $f(\cdot)$ is convex and
  nondecreasing in $[0, K]$, thus $|B'|\frac{f(|B|+1)-f(|B|)}{K}\left(1-\alpha\gamma \sum_{j_i\in A}\xi_i \right) \leq \frac{f(|B\cup B'|) - f(|B|)}{K}(1-\alpha\gamma \sum_{j_i\in A}\xi_i)$. 
  
  Now to continue with \labelcref{eq34}, one can verify that by setting $B = \emptyset, B' = \Omega$, the minimum of (\ref{eq34}) is achieved as $\gamma$, thus proving the submodularity ratio to be $\gamma$.

  Then for the \textbf{curvature}, for any $t\in T = A\cup B$,  we want 
  to lower bound $\frac{\rho_{t}(T\setminus \{t\}\cup T')}{\rho_{t}(T\setminus \{t\})}$. There are two cases: 
  
  {1)}  
   When $t = j_i \in A$, we have 
  \begin{flalign}\notag 
   \frac{\rho_{j_i}(T\setminus \{j_i\}\cup T')}{\rho_{j_i}(T\setminus \{j_i\})} & = \frac{\left[ 1 - \frac{\alpha\gamma}{K} f(|B\cup B'|)\right] \xi_i}{\left[ 1 - \frac{\alpha\gamma}{K} f(|B|)\right] \xi_i} \\ \label{subeq_41}
   & =  \frac{1 - \frac{\alpha\gamma}{K} f(|B\cup B'|)}{1 - \frac{\alpha\gamma}{K} f(|B|)}.
  \end{flalign}
  Since $f(\cdot)$ is convex and
    nondecreasing in $[0, K]$, 
  it is easy to see that the minimum of \labelcref{subeq_41} is achieved when $B=\emptyset, B' = \Omega$ as $1-\alpha$.

 {2)}   
 When $t = \omega_i\in B$, we have, 
  \begin{flalign}\notag 
    \frac{\rho_{\omega_i}(T\setminus \{\omega_i\}\cup T')}{\rho_{\omega_i}(T\setminus \{\omega_i\})}  
  & = \frac{\frac{f(|B\cup B'|) - f(|B\cup B'| - 1)}{K}\left[1-\alpha\gamma \sum_{i'\in A\cup A'}\xi_{i'}\right]}{\frac{f(|B|) - f(| B| - 1)}{K}\left[1-\alpha\gamma \sum_{i\in A}\xi_i\right]}\\\label{subeq_42}
  & \geq \frac{1-\alpha\gamma \sum_{i'\in A\cup A'}\xi_{i'}}{1-\alpha\gamma \sum_{i\in A}\xi_i}\\\label{subeq_43}
  &  = \frac{1- \alpha + \alpha - \alpha\gamma \sum_{i'\in A\cup A'}\xi_{i'}}{1-\alpha\gamma \sum_{i\in A}\xi_{i}} 
  \end{flalign}
  where \labelcref{subeq_42} is because $f(\cdot)$ is convex and
      nondecreasing in $[0, K]$. 
  
  Since $ \alpha - \alpha\gamma \sum_{i'\in A\cup A'}\xi_{i'} \geq 0$ and $-\alpha\gamma \sum_{i\in A}\xi_{i} \leq 0$, continuing with \labelcref{subeq_43} we have, 
  \begin{flalign}\notag 
  & \frac{\rho_{\omega_i}(T'\setminus \{\omega_i\}\cup T)}{\rho_{\omega_i}(T\setminus \{\omega_i\})} \geq 1-\alpha. 
  \end{flalign}
 
  The above two cases jointly prove that the objective in \labelcref{tight_fn} has 
  curvature  $\alpha$. 
  \end{proof}
  
\section{Existing Notions of Curvature and Submodularity Ratio}
  \label{app_classical_defs}

In this section we firstly discuss  existing
notions of curvature and submodularity ratio, then secondly
we present the relations to the notions in this paper. 
 
  \subsection{Classical Notions of  Curvature and Submodularity Ratio}
  {The curvature of submodular functions}  measures how close a submodular set function is to being modular, and has
  been used to prove improved theoretical results for constrained submodular 
  minimization and learning of  submodular functions \citep{iyer2013curvature}. Earlier, it has been used to tighten 
  bounds for submodular maximization subject to a cardinality 
  constraint \citep{conforti1984submodular} or a matroid constraint \citep{vondrak2010submodularity}.
  \begin{definition}[Curvature of submodular functions~\citep{conforti1984submodular,vondrak2010submodularity,iyer2013curvature}]
  The \emph{total curvature} $\kappa_F$ (which we term as $\alpha^{\text{\emph{total}}}$ in the main text) of a submodular function $F$ and the \emph{curvature} $\kappa_F(S)$ w.r.t.\ a set $S \subseteq \groundset$ are defined as,
  \begin{align}\notag 
     & \kappa_F \coloneqq 1 - \min_{j\in \groundset} \frac{\rho_j (\groundset \setminus \{j\})}{\rho_j(\emptyset)} \textnormal{ and}\\\notag 
       & \kappa_F(S) \coloneqq 1 - \min_{j\in S} \frac{\rho_j (S \setminus \{j\})}{\rho_j(\emptyset)},
  \end{align}
  respectively.
  Assume without loss of generality that $F(\{j\}) >0,  \forall j\in \groundset$. One can observe that $\kappa_F(S)\leq \kappa_F$. A modular
  function has curvature $\kappa_F = 0$, and a matroid rank function has maximal
  curvature $\kappa_F = 1$. 
  \citet{vondrak2010submodularity} also defines 
  the relaxed notion of curvature (which is called \emph{curvature with respect to the optimum}) to be the 
  smaller scalar $\bar \kappa_F(S)$ s.t,
   \begin{align}
   \rho_{T}(S) + \sum_{j\in S\cup T} \rho_j(S\cup T\setminus \{j \})  \geq (1- \bar \kappa_F(S)) \rho_{T}(\emptyset), \forall T\subseteq \groundset.  
   \end{align}

  \citet{iyer2013curvature} propose
  two new notions of curvature, which are,
  \begin{align}\notag 
  & \tilde{\kappa}_F(S): = 1 - \min_{T\subseteq \groundset} \frac{\rho_{T}(S) + \sum_{j\in S\cup T} \rho_j(S\cup T\setminus \{j \})}{\rho_{T}(\emptyset)},\\\notag 
  & \hat{\kappa}_F(S): = 1 - \frac{\sum_{j\in S}\rho_j(S\setminus \{j \})}{\sum_{j\in S}\rho_j(\emptyset)}.
  \end{align}
  \end{definition}
  \citet{iyer2013curvature} show that for submodular functions, it holds that $ \hat{\kappa}_F(S) \leq \kappa_F(S) \leq \tilde{\kappa}_F(S) \leq \kappa_F$.

  \paragraph{Submodularity ratio.}
  Informally, the submodularity ratio quantifies how close a set function is to being submodular~\citep{das2011submodular}.  
  \begin{definition}[Original submodularity ratio from \citet{das2011submodular}]\label{def_ratio_das}
%  Let $S, L\subseteq \groundset$ be two disjoint sets and 
  Let $F(\cdot)$ be a non-negative nondecreasing set function. 
  The submodularity ratio of a set $U$ w.r.t.\ an integer $k$ is given by, 
  \begin{flalign}\notag 
  \gamma_{U, k} \coloneqq \min_{L\subseteq U} \min_{L,S: L\cap S = \emptyset, |S|\leq k}\frac{\sum_{j\in L} \rho_j(S)}{\rho_L(S)}.
  \end{flalign}
  \end{definition}
  
  \subsection{Curvature of Non-submodular Functions and Relation  to Our Results} 
  \label{append_jan}
   
  \citet{sviridenko2015optimal} present a new notion of curvature
  for monotone set functions. 
We show how it is related to 
our notion of curvature in \cref{def_cur}. We also show
that our approximation factors using the  combination of  curvature and submodularity ratio
characterize the performance of \algname{Greedy} for solving  problem \labelcref{eq1} better. 

Specifically, for a nondecreasing function $F$,  \citet[Section 8]{sviridenko2015optimal} define the  curvature $c$ as
\begin{align}\label{app_def_c}
	1- c =  \min_{j\in \groundset} \min_{A, B\in \groundset\setminus \{j\}} \frac{\rho_j(A)}{\rho_j(B)}. 
\end{align}
 \citep[Theorem 8.1]{sviridenko2015optimal} show that  for maximizing a nondecreasing 
function with bounded curvature $c\in [0,1]$
under a matroid constraint, \algname{Greedy} enjoys an approximation guarantee 
of $(1-c)$, and it is tight  in terms
of the definition of $c$ in \labelcref{app_def_c}.   
The following remark discusses the relation to our definition of curvature. 
\begin{remark}\label{append_remark}
	For a nondecreasing function $F(\cdot)$, it holds: a) $c$ in \labelcref{app_def_c} is always 
	larger than the  notion of curvature $\alpha$ in \cref{def_cur}, i.e.,
	$c\geq \alpha$; b) For the \algname{Greedy} algorithm, there exists  a class
	of functions for which the  approximation guarantee characterized by
	$c$  (which is $1-c$) is strictly  smaller than 
	the  approximation guarantee
	characterized by the combination of $\alpha$ and $\gamma$ (which is ${\alpha^{-1}}(1-e^{-\alpha\gamma})$ according to \cref{thm_21}).
\end{remark}  
\begin{proof}[Proof of \cref{append_remark}]
\mbox{ }

	\textbf{a)} 
	Note that the  definition of curvature in \cref{def_cur} is
	equivalent to the smallest scalar $\alpha$ such that,
	\begin{align}\notag 
	\forall j\in \groundset,  \forall  B\subseteq A\in \groundset\setminus \{j\},\rho_j(A) \geq (1-\alpha)\rho_j(B). 
	\end{align}
	Now it is easy to see that $c\geq \alpha$. 
	
	\textbf{b)} Consider the class of functions
	in our tightness result in \labelcref{tight_fn}. 
	From  \cref{claim45} we know that  its curvature is $\alpha$ and submodularity
	ratio is $\gamma$. So its curvature $c$ in \labelcref{app_def_c} must be greater than or 
	equal to $\alpha$. Note that the  approximation guarantee characterized by
		$c$ is   $1-c \leq 1- \alpha$.  Taking $\alpha = 1$ in 
\labelcref{tight_fn},  the  approximation guarantee of 
	\citet{sviridenko2015optimal} is 0. While 
	our approximation guarantee is $\gamma$, 	
	 for any $\gamma\in (0, 1]$, our approximation 
	 guarantee is strictly  higher than $1-c$. 
\end{proof}

\subsection{Relation to  Notions  in This Work}

\begin{itemize}

\item There are two versions of submodularity ratio in this paper: $\gamma$ and $\gamma^G$,  $\gamma^G$ cannot be recovered from \cref{def_ratio_das}. Our theory can easily accommodate \cref{def_ratio_das}: our approximation guarantee in \cref{thm_21} holds for \cref{def_ratio_das} as long as $U$ contains $\opt$ and $k \geq  K$. One benefit of the  definition in this work (\cref{def:gen-submod-ratio}) is that it better handles subtleties in \cref{def_ratio_das} where the denominator could be $0$.

\item 
The  curvature in this work
is a  natural extension of the classical ones for monotone nondecreasing
submodular functions \citep{conforti1984submodular}. 

\item 
Note that  classical notions of curvature  
    measure how close a submodular set function is to being modular. The notions  of (generalized) curvature  in \cref{def_cur}  measures how close a set
    function is to being \emph{supermodular}. 
\item 
Our combinations of (generalized) curvature and submodularity
ratio gives   tight approximation
guarantees for \algname{Greedy}, and this 
combination
is  more expressive than the curvature by 	\citet{sviridenko2015optimal}, as shown in \cref{append_remark}. 
\end{itemize}

\section{Proofs for Bounding Parameters of  Applications}
  \label{sec_app_proof_app}

  \subsection{Proving  \cref{lemma_a_opt}}
  
  \begin{proof}[Proof of \cref{lemma_a_opt}]
  \mbox{}

Notice that in this subsection, the matrix  $\bmX_S = [\x_{v_1}, \ldots, \x_{v_s}]	 \in \R^{d\times |S|}$ is the submatrix
consisting the columns  of $\bmX$ indexed by the set $S$.

Our proof  
considers the spectral parameters of the matrix $\bmX_S \bmX_S^{\trans}$.
For brevity, let us write $\bmB = \bmLambda + \sigma^{-2}\bmX_S \bmX_S^{\trans}$.  $\bmB$ is a symmetric  positive definite
matrix, thus can be factorized as $\bmB = \bmP \bmD \bmP^{-1}$.

Let  the eigenvalues of $\bmX_S \bmX_S^{\trans}$ be  $\lambda_1(S)\geq \cdots \geq \lambda_d (S)\geq 0$, where we 
use the notation that $\lambda_i(S) \coloneqq \lambda_i(\bmX_S \bmX_S^{\trans}), \forall i\in [d]$. 
Then the eigenvalues of $\bmB$ are $\beta^2+\sigma^{-2}\lambda_i(S), i\in [d]$. 
One can  see that $\bmB^{-1} = \bmP \bmD^{-1} \bmP^{-1}$, 
and $\tr{\bmB^{-1}} = \tr{\bmD^{-1}} = \sum_{i=1}^{d} \frac{1}{\beta^2+\sigma^{-2}\lambda_i(S)}$. 

Let  the  singular values of $\bmX_S$ be  $\sigma_1(\bmX_S)\geq \cdots \geq \sigma_q(\bmX_S)$, where $q \leq \min\{d, |S|\}$. 
For notational simplicity, when $|S|< d$, we still use
the convention $\sigma_i(\bmX_S)=0, i = q+1,..., d$ 
to represent the zeros values. One has  $\sigma_i^2(\bmX_S) = \lambda_i(S), i = 1,..., d$.  
For notational  simplicity, we use $F(\cdot)$ to represent 
$F_A(\cdot)$ in the following. 

%  To make this paper self-contained, we state the 
% \textit{Cauchy interlacing property}  of singular values here:   
%let  $\bmA\in \R^{m\times n}$, its singular values to
%be $\sigma_1(\bmA)\leq \cdots \leq \sigma_q(\bmA)$, where $q=\min\{m, n\}$.
%Let $\bmB$  be a $(m-k)\times (n-l)$ submatrix of 
%$\bmA$, then  
%$\sigma_{i+k+l}(\bmA) \leq \sigma_i(\bmB) \leq \sigma_i(\bmA), \forall 1\leq i\leq q-(k+l)$. 
  
  \textbf{Monotonicity.} \quad 
  It can be easily seen that $F(\emptyset) = 0$. To prove that 
  $F(S)$ is monotone nondecreasing, one just needs to show that 
  $\forall \omega \in \groundset\setminus S$, it holds that $F(\{\omega\}\cup S)-F(S) \geq 0$. One can see that, 
  \begin{flalign}\notag 
   F(\{\omega\}\cup S)-F(S) &=\sum_{i=1}^{d}\frac{1}{\beta^2+\sigma^{-2}\sigma_i^2(\bmX_S)} - \sum_{j=1}^{d}\frac{1}{\beta^2+\sigma^{-2}\sigma_j^2(\bmX_{S\cup\{\omega\}})} \\\notag 
  &\geq 0 \quad (\text{Cauchy interlacing inequality of singular values}).
  \end{flalign}

  \paragraph{Bounding parameters.}
 
  Let us restate the \textit{assumption:} The data points are
  normalized, i.e., $\|\x_i\| = 1, \forall i\in \groundset$.
  Given this assumption, it holds that the spectral norm of the 
  data matrix $\|\bmX\| = \sigma_{\text{max}}(\bmX) \leq \sqrt{n}$, because of Weyl's inequality.

\textit{--Bounding the  submodularity ratio: }  
We need to lower bound $\frac{\sum_{\omega\in \Omega\backslash S}\rho_{\omega}(S)}{\rho_{\Omega}(S)} = \frac{\sum_{\omega\in \Omega\backslash S}F(\{\omega\}\cup S)-F(S)}{F(\Omega\cup S)-F(S)}$. 
  
  For the numerator, we have,
  \begin{flalign}\notag 
  \sum_{\omega\in \Omega\backslash S}F(\{\omega\}\cup S)-F(S)  
  &  = \sum_{\omega\in \Omega\backslash S}\left[\sum_{i=1}^{d}\frac{1}{\beta^2+\sigma^{-2}\sigma_i^2(\bmX_S)} - \sum_{j=1}^{d}\frac{1}{\beta^2+\sigma^{-2}\sigma_j^2(\bmX_{S\cup\{\omega\}})} \right]\\\notag 
  & =  \sum_{\omega\in \Omega\backslash S}\sum_{i=1}^{d}\frac{\sigma^{-2}[\sigma_i^2(\bmX_{S\cup\{\omega\}}) - \sigma_i^2(\bmX_S)]}{(\beta^2+\sigma^{-2}\sigma_i^2(\bmX_S))(\beta^2+\sigma^{-2}\sigma_i^2(\bmX_{S\cup\{\omega\}}))} \\\notag 
  & \geq (\beta^2+\sigma^{-2}\sigma_{\text{max}}^2(\bmX))^{-2} \sum_{\omega\in \Omega\backslash S}\sum_{i=1}^{d} \sigma^{-2}[\sigma_i^2(\bmX_{S\cup\{\omega\}}) - \sigma_i^2(\bmX_S)]\\\notag 
  & = (\beta^2+\sigma^{-2}\|\bmX\|^2)^{-2} \sum_{\omega\in \Omega\backslash S}\sum_{i=1}^{d} \sigma^{-2}[\lambda_i({S\cup\{\omega\}}) - \lambda_i(S)]\\\notag
  & = (\beta^2+\sigma^{-2} \|\bmX\|^2)^{-2} \sum_{\omega\in \Omega\backslash S} \sigma^{-2}[\tr{\bmX_{S\cup\{\omega\}} \bmX_{S\cup\{\omega\}}^\trans} - \tr{\bmX_{S} \bmX_{S}^\trans}]\\\notag 
  & = (\beta^2+\sigma^{-2}\|\bmX\|^2)^{-2} \sum_{\omega\in \Omega\backslash S} \sigma^{-2}[\tr{\bmX_{S} \bmX_{S}^\trans  + \x_{\omega}\x_ {\omega}^
  \trans} - \tr{\bmX_{S} \bmX_{S}^\trans}]\\\notag  
  & = (\beta^2+\sigma^{-2}\|\bmX\|^2)^{-2} \sum_{\omega\in \Omega\backslash S} \sigma^{-2}\tr{\x_{\omega}\x_ {\omega}^
    \trans}   \quad \text{ (linearity of the trace )}\\\notag  
    & = (\beta^2+\sigma^{-2}\|\bmX\|^2)^{-2} \sum_{\omega\in \Omega\backslash S}\sigma^{-2} \|\x_ {\omega}\|^2 \\\label{eq_45} 
    & =\sigma^{-2} (\beta^2+\sigma^{-2}\|\bmX\|^2)^{-2} |\Omega\setminus S| \quad \text{ (normalization of the data points)}
  \end{flalign}

%  Let $\Omega\setminus S =\{ \omega_1,\cdots, \omega_{|\Omega\setminus S|}\}$. 
  
  For the denominator, one has,
  \begin{flalign}\notag 
  F(\Omega\cup S)-F(S) & = \sum_{i=1}^{d}\frac{1}{\beta^2+\sigma^{-2}\sigma_i^2(\bmX_S)} - \sum_{j=1}^{d}\frac{1}{\beta^2+\sigma^{-2}\sigma_j^2(\bmX_{S\cup\Omega})} \\\notag 
 &  \leq  \sum_{i= d - |\Omega\setminus S|+ 1}^{d} \frac{1}{\beta^2+\sigma^{-2}\sigma_i^2(\bmX_S)} - \sum_{j=1}^{|\Omega\setminus S|}\frac{1}{\beta^2+\sigma^{-2}\sigma_j^2(\bmX_{S\cup\Omega})} \text{ (interlacing inequality of singular values)} \\\notag 
 & \leq |\Omega\setminus S| (\frac{1}{\beta^2} - \frac{1}{\beta^2+\sigma^{-2}\|\bmX\|^2})\\\label{subeq_init}
 & = |\Omega\setminus S|  \frac{\sigma^{-2}\|\bmX\|^2}{\beta^2(\beta^2+\sigma^{-2}\|\bmX\|^2)}. 
  \end{flalign}

  Combining \labelcref{eq_45} and \labelcref{subeq_init} yields,
  \begin{flalign}\notag 
   \frac{\sum_{\omega\in \Omega\backslash S}F(\{\omega\}\cup S)-F(S)}{F(\Omega\cup S)-F(S)} \geq 
  & \frac{|\Omega\setminus S| \sigma^{-2} (\beta^2+\sigma^{-2} \|\bmX\|^2)^{-2}}{|\Omega\setminus S|  \frac{\sigma^{-2}\|\bmX\|^2}{\beta^2(\beta^2+\sigma^{-2}\|\bmX\|^2)}}\\\notag 
  & = \frac{\beta^2}{\|\bmX\|^2(\beta^2+\sigma^{-2}\|\bmX\|^2)}.
  \end{flalign}
  
  \textit{--Bounding the curvature:} \quad 
  We want to lower bound $1 - \alpha$, which
  corresponds to lower bounding $  \frac{F(S \cup \Omega) -  F(S\setminus \{i\} \cup \Omega)}{F(S) - F(S\setminus \{i\})}  $.
%  \begin{flalign}\notag 
%   \frac{F(S \cup \Omega) -  F(S\setminus \{i\} \cup \Omega)}{F(S) - F(S\setminus \{i\})}  
%  \end{flalign}
  For the numerator, one has,
  \begin{flalign}\notag
   F(S \cup \Omega) -  F(S\setminus \{i\} \cup \Omega) 
  & = \sum_{i'=1}^{d}\frac{1}{\beta^2+\sigma^{-2}\sigma^2_{i'}(\bmX_ {S \setminus \{i\}  \cup \Omega})} - \sum_{j=1}^{d}\frac{1}{\beta^2+\sigma^{-2}\sigma_j^2(\bmX_ {S \cup \Omega})}\\ \label{curvature_numerator}
& \geq    \sigma^{-2} (\beta^2+\sigma^{-2}\|\bmX\|^2)^{-2}          \quad \text{ (similar derivation as in \labelcref{eq_45}) }.
  \end{flalign}
  
  For the denominator, one has (similar derivation as in \labelcref{subeq_init}),
  \begin{flalign}\notag 
   F(S) - F(S\setminus \{i\}) 
  & = \sum_{i'=1}^{d}\frac{1}{\beta^2+\sigma^{-2}\sigma^2_{i'}(\bmX_{S\setminus \{i\}})} - \sum_{j=1}^{d}\frac{1}{\beta^2+\sigma^{-2}\sigma_j^2(\bmX_S)}\\\notag 
  &\leq \frac{1}{\beta^2+\sigma^{-2}\sigma^2_{d}(\bmX_{S\setminus \{i\}})} - \frac{1}{\beta^2+\sigma^{-2}\sigma_1^2(\bmX_S)} \quad \text{ (Cauchy interlacing inequality)} \\\label{ieq_31}
  &\leq   \frac{\sigma^{-2}\|\bmX\|^2}{\beta^2(\beta^2+\sigma^{-2}\|\bmX\|^2)}.
  \end{flalign}
  
  Combining \labelcref{curvature_numerator,ieq_31} we  get,
  \begin{align}\notag 
    \frac{F(S \cup \Omega) -  F(S\setminus \{i\} \cup \Omega)}{F(S) - F(S\setminus \{i\})}  
   &  \geq \frac{\beta^2}{\|\bmX\|^2(\beta^2+\sigma^{-2}\|\bmX\|^2)}.
  \end{align}
  \end{proof}

\subsection{Proofs for Determinantal Functions of Square Submatrix}
    
    \begin{proof}[Proof of \cref{prop_ratio_determinant}]
    \mbox{ }
    
    Notice that in this subsection, the matrix $\bmSigma_S$
    is the square submatrix of $\bmSigma$, with both
    its rows and columns indexed by $S$.

    \textbf{a)} 
    We want to prove that $F(\cdot)$ is supermodular. Assume that
    $A\subseteq B\subseteq \groundset$ and $i\in \groundset\setminus B$, then
    \begin{flalign}\notag 
    \rho_i(A) &= \de{\bmI + \sigma^{-2}\bmSigma_{A\cup \{ i\}}} - \de{\bmI + \sigma^{-2}\bmSigma_A} \\\notag 
        & {=}\sum_{S\subseteq A \cup\{i\}} \de{(\sigma^{-2}\bmSigma)_{S}} -  \sum_{S\subseteq A} \de{(\sigma^{-2}\bmSigma)_{S}}  \quad \text{\citep[Theorem 2.1]{kulesza2012determinantal} }
     \\\notag 
    & {=}\sum_{S\subseteq A} \de{(\sigma^{-2}\bmSigma)_{S\cup\{i\}}}\\\notag 
    & \leq \sum_{S\subseteq B} \de{(\sigma^{-2}\bmSigma)_{S\cup\{i\}}} \quad \text{($\bmSigma$ is positive semidefinite)}\\\notag 
    &= \de{\bmI + \sigma^{-2}\bmSigma_{B\cup \{ i\}}} - \de{\bmI + \sigma^{-2}\bmSigma_B} \\\notag 
    & = \rho_i(B), 
    \end{flalign}
    which proves that $F(\cdot)$ is supermodular. 
    
    \textbf{b)}
     We want to lower bound $\frac{\sum_{\omega\in \Omega\backslash S}\rho_{\omega}(S)}{\rho_{\Omega}(S)} = \frac{\sum_{\omega\in \Omega\backslash S}F(\{\omega\}\cup S)-F(S)}{F(\Omega\cup S)-F(S)}$. 
     
     For the numerator, one has,
     \begin{flalign}\notag 
     \sum_{\omega\in \Omega\backslash S}F(\{\omega\}\cup S)-F(S)
      & =  \sum_{\omega\in \Omega\backslash S}\prod_{i=1}^{|S\cup \{\omega \}|}\lambda_i(\bmA_{S\cup \{\omega \}}) - \prod_{j=1}^{|S|}\lambda_j(\bmA_{S})\\\notag
     & =  \sum_{\omega\in \Omega\backslash S}\lambda_{|S\cup \{\omega \}|}(\bmA_{S\cup \{\omega \}}) \prod_{i=1}^{|S|}\lambda_i(\bmA_{S\cup \{\omega \}}) - \prod_{j=1}^{|S|}\lambda_j(\bmA_{S})\\\notag
    &  \geq \sum_{\omega\in \Omega\backslash S}\lambda_{|S\cup \{\omega \}|}(\bmA_{S\cup \{\omega \}}) \prod_{i=1}^{|S|}\lambda_i(\bmA_{S}) - \prod_{j=1}^{|S|}\lambda_j(\bmA_{S})  \quad \text{ (Cauchy interlacing inequality)} \\\label{eq52}
    &  = \sum_{\omega\in \Omega\backslash S}(\lambda_{|S\cup \{\omega \}|}(\bmA_{S\cup \{\omega \}}) -1) \prod_{i=1}^{|S|}\lambda_i(\bmA_{S}).
     \end{flalign}
     
    For the denonimator, it holds, 
    \begin{flalign}\notag 
    F(\Omega\cup S)- F(S) 
    & = \prod_{i=|\Omega\setminus S|}^{|\Omega\cup S|}\lambda_i(\bmA_{\Omega\cup S})\prod_{j=1}^{|\Omega\setminus S|}\lambda_j(\bmA_{\Omega\cup S}) - \prod_{i=1}^{|S|}\lambda_i(\bmA_{S})\\\label{eq54}
    & \leq  \left(\prod_{j=1}^{|\Omega\setminus S|}\lambda_j(\bmA_{S\cup \Omega}) -1\right) \prod_{i=1}^{|S|}\lambda_i(\bmA_{S}) \quad \text{(Cauchy interlacing inequality)}.
    \end{flalign}
    
    Combining \labelcref{eq52,eq54} gives,  
    \begin{flalign}\notag 
     \frac{\sum_{\omega\in \Omega\backslash S}F(\{\omega\}\cup S)-F(S)}{F(\Omega\cup S)-F(S)} 
     & \geq \frac{\sum_{\omega\in \Omega\backslash S}(\lambda_{|S\cup \{\omega \}|}(\bmA_{S\cup \{\omega \}}) -1) \prod_{i=1}^{|S|}\lambda_i(\bmA_{S})}{ \left(\prod_{j=1}^{|\Omega\setminus S|}\lambda_j(\bmA_{S\cup \Omega}) -1\right) \prod_{i=1}^{|S|}\lambda_i(\bmA_{S})}\\\notag
    & = \frac{\sum_{\omega\in \Omega\backslash S}(\lambda_{|S\cup \{\omega \}|}(\bmA_{S\cup \{\omega \}}) -1)}{ \left(\prod_{j=1}^{|\Omega\setminus S|}\lambda_j(\bmA_{S\cup \Omega}) -1\right)}\\\notag
    & \geq \frac{K(\lambda_n - 1)}{\prod_{j=1}^{K}\lambda_j -1},
    \end{flalign}
    where the last inequality comes from that $|\Omega\setminus S|\leq K$. 
    \end{proof}
  
 \subsection{LP with Combinatorial Constraints}
  \label{app_subsec_lp}

\subsubsection{Two  examples where $F(S)$ is non-submodular}

1), Considering the following LP: 
\begin{align} 
\begin{array}{r l l l l}
\max &  4x_1+ & x_2 + & 4x_3\\
\text{s.t.} & 2x_1+ & x_2 &  & \leq 2\\
				& 			& x_2+ & 2x_3 & \leq 2\\
\end{array}\\\notag 
x_1, x_2, x_3 \geq 0. 
\end{align}
For this LP, one can easily see that $F(\{1,2\}) = 4, F(\{2\})= 2, F(\{1,2,3\})=8, F(\{2,3\})=4$, thus $F(\{1,2\}) - F(\{2\}) <  F(\{1,2,3\})-F(\{2,3\})$, which shows $F$ is non-submodular.

2), Considering the following LP: 
\begin{align} 
\begin{array}{r l l l l}
\max &  10x_1+ & 12 x_2 + & 12x_3\\
\text{s.t.} & x_1+ &2 x_2 + & 2x_3  & \leq 20\\
				& 	2x_1+		& x_2+ & 2x_3 & \leq 20\\
				& 2x_2 + & 2x_2 +  & x_3   & \leq 20
\end{array}\\\notag 
x_1, x_2, x_3 \geq 0. 
\end{align}
For this LP, one can  see that $F(\{1,2\}) = 120, F(\{2\})= 120, F(\{1,2,3\})=136, F(\{2,3\})=120$, thus $F(\{1,2\}) - F(\{2\}) <  F(\{1,2,3\})-F(\{2,3\})$. But this one has \textit{degenerate} basic feasible
solutions.

\subsubsection{Proving \cref{lemma_lp_ratio}}
\label{app_subsub_lp}

To prove \cref{lemma_lp_ratio}, we first need to 
present the setup. 
The LP corresponding to
$F(S)$ is, 
\begin{align} 
\begin{array}{r l l l l}
&\max &  \dtp{\d_S}{\x_S}\\
(LP_S) & \text{s.t.} & \bmA_S \x_S \leq \b
\end{array}\\\notag 
\x_S \geq 0. 
\end{align}
where the columns of  $\bmA_S\in \R_+^{m\times |S|}$ are  the columns
of $\bmA$ indexed by the set $S$. $\x_S$ (respectively, $\d_S$) is the subvector of $\x$ (respectively,  $\d$) indexed by $S$.  To apply the optimality 
condition of a LP in the standard form, let us change $(LP_S)$ to be the following standard LP by introducing the slack variable $\bmxi\in \R^m$, 
\begin{align} 
\begin{array}{r l l l l}
& -\min &  \dtp{\c_S}{\x_S}\\
(LP_S^*) & \text{s.t.} & \bmA_S \x_S + \bmI_m \bmxi = \b
\end{array}\\\notag 
\x_S \geq 0, \bmxi \geq 0. 
\end{align}
where $\c_S := -\d_S$.  Let us denote $\bar \bmA :=[\bmA_S, \bmI_m] \in \R^{m\times (|S|+m)}$, $\bar \x:=[\x_S^\trans, \bmxi^\trans]^\trans$.

Let $(\x^{(S)}, \bmxi^{(S)})$ denote the optimal solution of $(LP^*_S)$.
% and $\x^{(A)}_B\in \R^n$ denote $\x^{(A)}$ with all the entries  on the set $\groundset\setminus B$ zeroing out. 
%Formally, $\x^{(S)} = F(S)= \argmax_{\spt{\x}\subseteq S,  \; \x\in \P} \dtp{\d}{\x}$. 
The corresponding basis of of $(LP^*_S)$ is $B^{(S)}$, which is a subset of $\groundset \cup \{\xi_1, \cdots, \xi_m \}$, and $|B^{(S)}| = m$.

According to \citet[Chapter 3.1]{bertsimas1997introduction}, the \textit{optimality condition} for  $(LP^*_S)$ is: Given a basic feasible solution $(\x, \bmxi)$ with the basis as $B$, the reduced cost is $\bar c_j = c_j - \c_B^\trans \bar \bmA_B^{-1}\bar\bmA_{\cdot j}$. 1) If $(\x, \bmxi)$ is optimal and non-degenerate, then $\bar c_j \geq 0, \forall j$; 2) If $\bar c_j \geq 0, \forall j$,
then $(\x, \bmxi)$ is optimal. 
  
\begin{proof}[Proof of \cref{lemma_lp_ratio}]
First of all, let us detail the non-degenerancy assumption.

\textbf{Non-degenerancy assumption:}
The basic feasible solutions of the correpsonding LP in standard 
form ($LP^*_S$) is non-degenerate $\forall S\subseteq \groundset$.

\textbf{a)}
It is easy to see that $F(\emptyset) = 0$, and $F(S)$ is nondecreasing. 

\textbf{b)}
For the submodularity ratio, we want to lower bound $\frac{\sum_{\omega\in \Omega\backslash S}\rho_{\omega}(S)}{\rho_{\Omega}(S)}$.
There could be in total four situations:

1)   $\sum_{\omega\in \Omega\backslash S} \rho_{\omega}(S) = 0$ but $\rho_{\Omega}(S) >0$. We will prove that  this situation  cannot happen, or in the other words, 
$\sum_{\omega\in \Omega\backslash S}F(\{\omega\}\cup S)-F(S) = 0$ implies that $F(\Omega\cup S)-F(S) =0$ as well. 
 
 First of all, since $F(S)$ is nondecreasing, so $F(\{\omega\}\cup S)-F(S)=0, \forall \omega$. We know that $(\x^{(S)}, \bmxi^{(S)})$ is the optimal solution of $(LP^*_S)$, and $(\x^{(S)}, \bmxi^{(S)})$ is a basic feasible 
 solution of $(LP^*_{S\cup\{\omega \}})$, so $(\x^{(S)}, \bmxi^{(S)})$ is also the optimal solution of $(LP^*_{S\cup\{\omega \}})$. 
 Since $(LP^*_{S\cup\{\omega \}})$ is non-degenerate, according to the 
 optimality condition, the reduced cost of $x_{\omega}$: $\bar c_{\omega}$
 must be greater than or equal zero. 
 
 Now we know that  $\bar c_{\omega} \geq 0, \forall \omega\in \Omega\backslash S$, and $(\x^{(S)}, \bmxi^{(S)})$ is a basic feasible 
 solution of $(LP^*_{S\cup\Omega})$ as well, again using the 
 optimality condition, we know that $(\x^{(S)}, \bmxi^{(S)})$ is optimal 
 for $(LP^*_{S\cup\{\Omega \}})$. So $F(\Omega\cup S)-F(S) =0$. 
 
2)  $\sum_{\omega\in \Omega\backslash S}\rho_{\omega}(S) = 0$ 
and $\rho_{\Omega}(S)=0$. The submodularity ratio is $1$ in this situation.

3) $\sum_{\omega\in \Omega\backslash S}\rho_{\omega}(S) > 0$ 
and $\rho_{\Omega}(S) =0$.  This can be ignored since we want a lower bound. 

4) $\sum_{\omega\in \Omega\backslash S}\rho_{\omega}(S) > 0$ 
and $\rho_{\Omega}(S) >0$. This situation gives the lower bound: 
\begin{align}\notag 
\frac{\sum_{\omega\in \Omega\backslash S}\rho_{\omega}(S)}{\rho_{\Omega}(S)} & \geq \frac{\max_{\omega\in \Omega\backslash S}\rho_{\omega}(S)}{F(\groundset)}\\\notag
& \geq  \frac{\min_{S\subseteq \groundset,  \omega\in \groundset\backslash S, \rho_{\omega}(S)>0}\rho_{\omega}(S)}{F(\groundset)}\\\notag
& =: \gamma_0 > 0.
\end{align}
\end{proof}

%\subsection{More on feature selection}
%
%
%
%\paragraph{@Other possible assumptions.}
%$f$ has $l$-Lipschitz continuous function value:
%\begin{flalign}
%|f(\x)-f(\y)|\leq l\|\x-\y\| , \forall \x, \y\in \dom f
%\end{flalign}
%which implies that $\|\nabla f(x)\| \leq l, \forall \x$.
%
%$f$ has $M$-Lipschitz continuous Hessian,
%\begin{flalign}
%\|\nabla^2 f(\x) - \nabla^2 f(\y)\| \leq M\|\x-\y\|, \forall \x, \y\in \dom f
%\end{flalign}
%which implies that,
%\begin{flalign}\notag 
%& |f(\y) - f(\x) -\dtp{\nabla f(\x)}{\y-\x} -\frac{1}{2}\dtp{\y-\x}{\nabla^2f(\x)(\y-\x)}|\\
%& \leq \frac{M}{6}\|\y-\x\|^3.
%\end{flalign}
%\paragraph{Bounding curvature.}
%
%Now let us try to upper bound the curvature of $F(S)$, which is equivalent to lower bounding $1-\alpha = \min_{\Omega: |\Omega| = K, S: |S|<K}\min_{i\in S \backslash \Omega} \frac{\rho_{i}(S\setminus \{i\} \cup \Omega)}{\rho_{i}(S\setminus \{i\})} $.  So we consider, 
%\begin{flalign}
%\min_{\Omega, S, i\in S \backslash \Omega}\frac{F(S \cup \Omega) -  F(S\setminus \{i\} \cup \Omega)}{F(S) - F(S\setminus \{i\})} 
%\end{flalign}

\section{Details about SDP Formulation of Bayesian A-optimality Objective}
\label{sdp_a_opt}
  
The SDP formulation used in this paper is consistent with that from 
    \citet[Chapter 7.5]{boyd2004convex} and \citet{krause2008near}. To make this work self-contained, we present  the details here.

  Firstly, maximizing the Bayesian A-optimality objective is equivalent
  to, 
  \begin{flalign}\label{eq_bayesian_a}
  \min_{S\subseteq \groundset, |S| \leq K} \tr{(\bmLambda + \sigma^{-2}\bmX_S \bmX_S^{\trans} )^{-1}}
  \end{flalign}
By introducing binary variables $m_j, j\in [n]$, \labelcref{eq_bayesian_a}
is equivalent to,
\begin{align}
& \min \tr{(\bmLambda + \sigma^{-2}\sum_{j=1}^{n}m_j \x_j \x_j^{\trans} )^{-1}}\\\notag 
& \text{ s.t. } m_j \in \{0, 1 \}, j\in [n], m_1 + \cdots + m_n \leq K
\end{align}
A proper  relaxation is (relaxing the variables $\lambda_j = m_j/K, j
\in [n]$), 
\begin{align}\label{a_opt_relaxation}
& \min \tr{(\bmLambda + \sigma^{-2}\sum_{j=1}^{n}\lambda_j \x_j \x_j^{\trans} )^{-1}}\\\notag 
& \text{ s.t. } \bmlambda \in \R^n_+,   \mathbf{1}^\trans \bmlambda = 1.
\end{align}
 According to the Schur complement lemma, the relaxed formulation \labelcref{a_opt_relaxation} is equivalent to the following SDP
 problem,
\begin{align}\notag 
 & \min_{\u\in \R^d} \mathbf{1}^\trans \u \\
\text{s.t. }  &
\begin{bmatrix}
  \bmLambda + \sigma^{-2}\sum_{j=1}^{n}\lambda_j \x_j \x_j^{\trans}     & \e_k \\\notag 
 \e_k^\trans &    u_k
\end{bmatrix}
\succeq 0, \quad  k = 1,\cdots,  d  \quad (\text{SDP})  \\\notag
 & \bmlambda \in \R^n_+,   \mathbf{1}^\trans \bmlambda = 1,
\end{align}
where $\e_k\in \R^d$ is the $k^\text{th}$ standard basis vector. 
According to \citet{krause2008near}, after solving the (SDP) problem we sort the
entries of $\bmlambda$ in descending order, and 
select the largest $K$ coordinates as the indices
of the $K$ elements to be selected.

 \section{Proofs and Details in Related Work  (\cref{sec_related})}
 \label{sec_related_work}
 
 \begin{remark}\label{counter_ex}
 For a set function $F(\cdot)$: 
 a)  Its   submodularity ratio  $\gamma$ is lower-bounded away from 0 and its curvature  $\alpha $ is  upper-bounded 
 away from 1 does not imply that it is weakly submodular;
 b) $F(\cdot)$ is weakly submodular  does not 
 imply that its   submodularity ratio  $\gamma$ is lower-bounded away from 0 and its curvature  $\alpha $ is  upper-bounded 
  away from 1.
 \end{remark}
 
 \begin{proof}[Proof of \cref{counter_ex}]\mbox{ }
 	
 	For argument a): Let $F(S) := |S|^4, S\subseteq \groundset$, which is a supermodular 
 	function, so the curvature is $0$ (upper-bounded away from 1). The submodualrity
 	ratio can be lower bounded by $n^{-3}$. But it is not weakly submodular according to Proposition 3.11 in  \citet{borodin2014weakly}.
 	
 	For argument b):  Let us take a minimum cardinality function
 	with $k=2$, i.e., $F(S) = B>0$ iff. $|S| \geq 2$, otherwise $F(S) =0$. 
 	According to Proposition 3.5 in \citet{borodin2014weakly}, it is 
 	weakly submodular, but it is easy to see that its submodualrity ratio is
 	$0$. 
 \end{proof}
 
 \iffalse
 \paragraph{More on supermodularity degree.} It is introduced by  \cite{feige2013welfare} to quantify to which extent
 a function exhibits supermodular properties.  
 For a set function $F$, the supermodular dependency set of an item $v$, $\dep^+(v)$, is the set of all items $v'$ in $\groundset$ so that there exists $S\subseteq \groundset\setminus \{v\}$ such that $\rho_v(S) > \rho_v(S\setminus \{v'\})$. Formally,  
 $$\dep^+(v) = \{v' \mid \exists S\subseteq \groundset\setminus \{v\}  , \text{ s.t. } \rho_v(S) > \rho_v(S\setminus \{v'\})\}.$$
 The supermodularity degree of $F$ is defined as $\max_{v\in \groundset}|\dep^+(v)|$, i.e., the largest cardinality of the supermodular dependency set of any item.

 For a specific  ground set  $\groundset = \{v_1, v_2, b\}$, define $F(S) = b>0$ if $S\supseteq \{v_1, v_2\}$, 
 otherwise $F(S) = 0$. One can observe that its supermodularity degree
 is $1$.  But its submodularity ratio can be 0. 
 Furthermore, any submodular function has supermodularity degree as $0$, but it has bounded submodularity ratio. 
 \fi

 \paragraph{More on submodularity index.} 
  It  is defined as (equivalent to that in \citet{NIPS2016_6384}): 
 \begin{align}\notag 
 \min_{\Omega, S\subseteq \groundset}\min_{|\Omega\setminus S|\leq K} \Big( {\sum_{\omega\in \Omega\backslash S}\rho_{\omega}(S)} - {\rho_{\Omega}(S)} \Big).
 \end{align}
 
  \iffalse
 which is  closely related to the submodularity ratio by \citet{das2011submodular}.
 
 \iffalse 
 It is noteworthy that \citet{NIPS2016_6384} implicitly assume  $|\Omega\setminus S| \geq 2$,
  so that the  submodularity index can be  strictly greater than zero (equivalently, the submodularity ratio can be  greater than one). Based on which they claim that the deterministic greedy algorithm (i.e., \algname{Greedy})  gets a stronger
    bound than $1-1/e$  (Corollary 1 therein), 
 for maximizing  monotone submoduar functions s.t. a $K$-cardinality constraint.

 However this implicit assumption may not   hold in general 
 for \algname{Greedy}, since in the last step of \algname{Greedy},  $|\Omega\setminus S|$ could be exactly one. To see this our tightness result in \cref{sec_tightness}
 can serve as a counter example: For the set function $F(T)$ in \labelcref{tight_fn},
  setting  $\gamma=\alpha=1$, now it is a monotone nondecreasing
 submodular function.  However, the approximation ratio of
 \algname{Greedy} for maximizing $F(T)$ s.t. a $K$-cardinality constraint  is precisely $1-1/e$. 
% which contradicts \citet[Corollary 1 ]{NIPS2016_6384}. 
 \fi 
 
 Note that our analysis of the \algname{Greedy} algorithm, which  
 considers a novel combination of  the (generalized) submodular ratio and 
 curvature, is  different from the classical analysis. Furthermore,
 it provides \textit{stronger} bounds
 for the maximization of monotone submodular functions as long as the 
 the curvature is upper-bounded away from 1. 
 \fi

\section{More Applications}
\label{app_more_apps}

\subsection{Subset Selection Using the  $R^2$ Objective}
\label{app_subsec_r2_intro}
Subset selection aims to estimate a predictor variable $Z$ using 
linear regression on a small subset from the set of 
observation variables $\groundset=\{X_1,..., X_n \}$.
Let $\bmC$ to be the covariance matrix among the observation variables $\{X_1,..., X_n \}$. We use $\b$ to denote 
the covariances between $Z$ and the $X_i$, with entries $b_i = \text{Cov}(Z, X_i)$.
Assuming there are $m$ observations, let us arrange the data of all the observation variables to be a 
design matrix $\bmX\in \R^{m\times n}$, with each column representing the observations of one variable. 
Given a budget parameter $K$, subset selection tries to find a set $S\subseteq \groundset$
of at most  $K$ elements, and a linear predictor $Z' = \sum_{i\in S}\alpha_i X_i = \bmX_{\cdot S}\bmalpha_S$, in order to maximize the squared multiple corrleation
% \citep{johnson2002applied}, 
$R_{Z, S} = \frac{\text{Var}(Z) - \E[(Z-Z')^2]}{\text{Var}(Z)}$, it measures
the fraction of variance of $Z$ explained by variables in $S$. Assume $Z$ is normalized to have variance 1, and it is well-known that the optimal 
regression coefficients are $\bmalpha_S = (\bmC_S)^{-1}\b_S$, so the 
 $R^2$ objective can be formulated as,
\begin{flalign}\label{eq_r2}
F(S) := R^2_{Z, S} = \b_S^{\trans}(\bmC_S)^{-1}\b_S, S\subseteq \groundset.
\end{flalign} 
%Assume throughout the work that $\bmC_S$ is non-singular. 
\citet{das2011submodular} show that the submodularity ratio of $F$ in \labelcref{eq_r2} can
be lower bounded by $\lambda_{\min}(\bmC)$, which is the smallest 
eigenvalue of $\bmC$. 
The  theoretical results in this work suggests that the approximation guarantees for maximizing $F$ in \labelcref{eq_r2}  can be further improved by analyzing the curvature parameters. The experimental results in \cref{subsec_r2}
   demonstrates that it is promising to upper bound the curvature parameters  of \labelcref{eq_r2} (possibly  with regular  assumptions) .

\subsection{Sparse Modeling with  Strongly Convex Loss Functions} 

%Feature selection \citep{guyon2003introduction} is a fundamental task in machine learning
%and data mining.  

Sparse modeling aims to build a model with a
small subset of at most $K$ features, out of
in total $n$ features. Let $f(\x): \R^n \mapsto \R$  to be 
the loss function, the corresponding objective is,
\begin{align}\notag 
\min f(\x) \text{ s.t. } |\spt{\x}| \leq K. 
\end{align}
Assume  $f(\x)$ is $m$-strongly convex and has Lipschitz continuous gradient with parameter $L$, which is equilavent to say that $g(\x):= -f(\x)$ is $m$-strongly concave and has $L$-Lipschitz continuous gradient. Then for all $\x, \y\in \dom{f}$ it holds, 
\begin{flalign}\label{eq_convex_smooth}
\frac{m}{2}\|\y - \x\|^2 \leq -g(\y) + g(\x) + \dtp{\nabla g(\x)}{\y-\x}\leq \frac{L}{2}\|\y - \x\|^2.
\end{flalign}
%Let $\x^{(A)}\in \R^n$ denote the optimal solution with maximal support as the set $A$, and $\x^{(A)}_B\in \R^n$ denote $\x^{(A)}$ with all the entries  on the set $\groundset\setminus B$ zeroing out. 
%Formally, $\x^{(A)} = \argmax_{\spt{\x}\subseteq A}g(\x)$.
In  solving 
this problem, the \algname{Greedy} algorithm maximizes the corresponding  auxiliary set function, 
\begin{align}\label{auxili_sparse}
F(S) : =\max_{\spt{\x}\subseteq S}g(\x), \; S\subseteq [n]
\end{align}
\citet{elenberg2016restricted} analyzed the approximation guarantees of \algname{Greedy} by bounding the submodularity ratio of $F(S)$.  Specifically, 
\begin{lemma}[Paraphrasing Theorem 1 in  \citet{elenberg2016restricted}]\label{lem_ratio_general}
The submodularity ratio of $F(S)$ in \labelcref{auxili_sparse} is lower bounded by $\frac{m}{L}$.
\end{lemma}
By further bounding the curvature parameters of the 
auxiliary set function in \labelcref{auxili_sparse}, one 
can get improved approximation guarantees according
to our theoretical findings.

\subsection{Optimal Budget Allocation with Combinatorial Constraints}

Optimal budget allocation \citep{soma2014optimal} 
is  a  special case of the influence maximization
problem, it aims to  distribute the budget (e.g., space of an inline advertisement, or time for a TV
advertisement) among the customers, and
to maximize the expected influence on the potential customers.
A concrete application is for the \emph{search
marketing advertiser bidding} task, in which vendors bid for the
right to appear alongside the results of different search
keywords. 
Let $x^i_s\in \R_+$ to be  the volume of advertising space allocated to the
advertiser $i$ to show his ad alongside query keyword $s$. 
\citet{bian2016guaranteed} present continuous DR-submodular objectives to model this problem with continuous assignments. 

The search 
engine company (e.g., Google and Yahoo) needs to distribute the budget (ad space) to all vendors
to maximize their influence on the customers,
while respecting various continuous and \emph{combinatorial} constraints. 
For the continuous constraints, for instance,  each vendor has a
specified budget limit for advertising, and the ad space associated
with each search keyword can not be too large. 
These continuous constraints can be
formulated as a convex set  $\P$. For \emph{combinatorial 
constraints}, each vendor needs to obey the Internet 
regulations of sensitive search keywords in his country, 
so the search engine company can only choose a subset of ``legal"
keywords for a specific vendor. The combinatorial constraints
can be arranged as a matroid  $\M = (\groundset, \I)$.
Hence  the problem in general  can be formulated as, $$\max_{\x\in \P \text{ and } \spt{\x}\in \I} g(\x),$$
where
$g(\x)$ is the total influence modeled by a DR-submodular function.  For one of its possible forms, one can refer to
\citet{bian2016guaranteed}. The \algname{Greedy} algorithm
solves this problem by maximizing the following auxiliary
set function $F(S)$ while respecting the combinatorial constraints,
\begin{align}\label{objective_inf}
\max_{S\in \I}F(S), \text{ where } F(S):= \max_{\spt{\x}\subseteq S, \x\in \P} g(\x).
\end{align}
By studying the submodularity ratio and curvature parameters
of $F(S)$ in \labelcref{objective_inf}, one could obtain 
theoretical guarantees of the \algname{Greedy} algorithm according to \cref{thm_21} in this work.

%
%
%Given is a bipartite graph
%$(S,T; W)$, where $S$ and $T$ are collections of advertising channels
%and customers, respectively.
%The edge weights, $p_{st}\in W$,  represent the influence probabilities.
%The goal is to
%.  
%
%
%
%The total influence of customer $t$ from all
%channels can be modeled by a proper continuous function \citep{bian2016guaranteed}
%$I_t(\x)$, e.g., $I_t(\x) = 1- \prod_{(s, t)\in W} \left(1-p_{st} \right)^{x_s}$ where $\x\in \R^S_+$ is the budget assignment among the advertising channels.
%For a set of $k$ advertisers, let $\x^i\in \R^S_+$ to be the budget assignment for
%advertiser $i$, and $\x:= [\x^1,\cdots, \x^k]$ denote the assignments for all the advertisers.    The overall objective is,
%\begin{flalign}\notag 
%g(\x)= \sum_{i=1}^k \alpha_i f(\x^i) ~\text{ with }~ f(\x^i)
%:=\sum_{t\in T} I_t(\x^i), \\\notag 
%0\leq \x^i\leq \bar u^i , \forall i = 1,\cdots, k
%\end{flalign}
%
\setkeys{Gin}{width=0.31\textwidth}
\begin{figure}[h!]
\center 
\subfloat[A-optimality objective \label{subfig_a_1}]{
\includegraphics[]{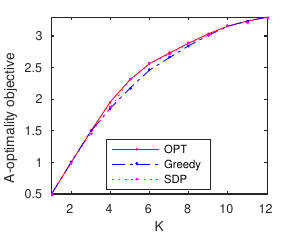}}
\subfloat[Greedy submodularity ratio and curvature \label{subfig_a_2}]{
\includegraphics[]{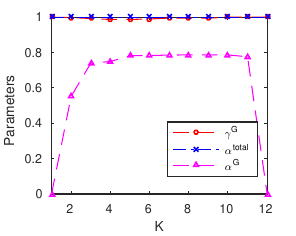}}
\subfloat[Approximation bounds \label{subfig_a_3}]{
\includegraphics[]{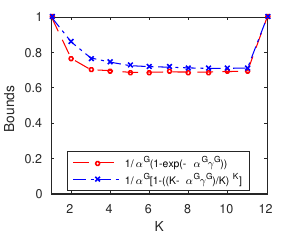}}
\caption{Function value, parameters and approximation bounds of experimental design on synthetic data. Correlation:  $0.5$}
\label{fig_syn_a}
\end{figure}

\section{More Experimental Results}
\label{more_exps}

\subsection{Bayesian A-optimality Experiments}

We put the results on a  randomly generated
dataset, to illustrate what does the proved bounds
looks like. 
In the synthetic experiments  we generate 
random observations from a multivariate Gaussian distribution with 
correlation  $0.5$. 
  \cref{fig_syn_a} shows the results (function value, parameters and approximation bounds) for one randomly generated data set with $d=6$ features and $n=12$ observations.  Specifically,   \cref{subfig_a_3}
   traces the two approximation bounds from \cref{thm_21} (and \cref{lem_34}): one curve shows
   the constant-factor bound ${\alpha^{-1}}(1-e^{-\alpha\gamma})$ and the other  the $K$-dependent 
   bound $ \frac{1}{\alpha} \left[1- \left(\frac{K-\alpha\gamma}{K}\right)^K\right]$. We observe  that both bounds give reasonable   predictions 
  of the performance of  \algname{Greedy}.

%\vspace{-0.3cm}
 \subsection{Subset Selection Using the $R^2$ Objective}
 \label{subsec_r2}

\setkeys{Gin}{width=0.32\textwidth}
\begin{wrapfigure}{l}{0.65\textwidth}
\vspace{-0.2cm}
   \center 
       \addtocounter{subfigure}{-2}
   \subfloat{
   \includegraphics[]{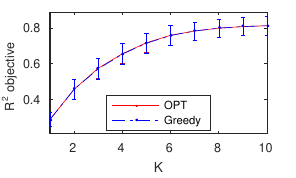}}
   \subfloat{
   \includegraphics[]{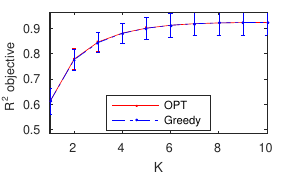}}\\
   \vspace{-.698cm}
   \subfloat[Correlation: 0.05 \label{fig_r2_syn1}]{
   \includegraphics[]{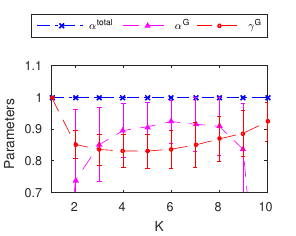}}
    \subfloat[Correlation: 0.5 \label{fig_r2_syn2}]{
      \includegraphics[]{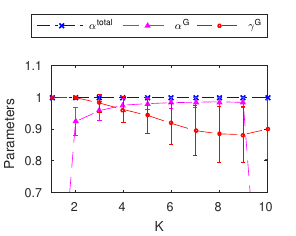}}
   \caption{Results  for $R^2$ objective on synthetic data.}
   \label{fig_r2_syn}
   \end{wrapfigure}
% \vspace{-0.4cm}

 For details on this task please refer to 
 \citet{das2011submodular} or \cref{app_subsec_r2_intro}.
% Function values and parameters
% for real-world data are presented  in \cref{app_subsec_r2}. 
 We did  synthetic experiments to illustrate that our theory can give a refined 
 explanation of the performance of  \algname{Greedy}.  
We generate 
  random observations from a multivariate standard Gaussian distribution with different 
  correlations. We used $n=10$ features and $m=100$ observations.  
The target regression coefficients $\bmalpha\in \R^n$ were  generated  as 
  a random vector with uniformly distributed entries in $[0, 1]$. Standard Gaussian
  noise was added to generate the observation
  of predictor variable $Z$.  
 The results are shown in \cref{fig_r2_syn}, with 
 first column showing the results with correlation as 0.05,
 the second column with correlation as 0.5. 
 One can see that the mean of the greedy curvature and 
 submodularity ratio take values in $(0, 1)$, which
 can be used to give improved approximation bounds
 for \algname{Greedy}. 
%\setkeys{Gin}{width=0.23\textwidth}
%\begin{figure}[htbp]
%   \center 
%       \addtocounter{subfigure}{-2}
%   \subfloat{
%   \includegraphics[]{Syntheticn10m100c0-05sig1.pdf}}
%   \subfloat{
%   \includegraphics[]{Syntheticn10m100c0-5sig1.pdf}}\\
%   \vspace{-.698cm}
%   \subfloat[Correlation: 0.05 \label{fig_r2_syn1}]{
%   \includegraphics[]{Syntheticratio_cur_param_n10m100c0-05sig1.pdf}}
%    \subfloat[Correlation: 0.5 \label{fig_r2_syn2}]{
%      \includegraphics[]{Syntheticratio_cur_param_n10m100c0-5sig1.pdf}}
%   \caption{Results  for $R^2$ objective on synthetic data.}
%   \label{fig_r2_syn}
%   \end{figure}
%% \vspace{-0.4cm}